\documentclass[a4paper,11pt]{article}
\pdfoutput=1 

\usepackage{jheppub} 

\usepackage[T1]{fontenc} 
\date{\today}

\usepackage{amsmath}
\usepackage{amsfonts}
\usepackage{graphicx}
\usepackage[english]{babel}
\usepackage{amssymb}
\usepackage{enumitem}
\begin{document}

\newcommand\fract[2]{{\textstyle\frac{#1}{#2}}}
\newcommand\halft{\fract{1}{2}}

\title{Boundary scattering in the $\phi^4$ model}

\author{Patrick~Dorey$^{a}$, Aliaksei~Halavanau$^{b,c}$,
James Mercer$^{a}$,
Tomasz~Romanczukiewicz$^{d}$ and
Yasha~Shnir$^{b,e,f}
$}
\affiliation{$^{a}$Department of Mathematical Sciences, Durham
University, UK\\
$^{b}$Department of Theoretical Physics and Astrophysics,
Belarusian State University,\\ Minsk,  Belarus \\
$^{c}$Department of Physics, Northern Illinois University,
USA\\
$^{d}$Institute of Physics, Jagiellonian University, Krakow,
Poland \\
$^{e}$BLTP, JINR, Dubna, Russia
\\ $^{f}$Institute of  Physics, Oldenburg University,
Germany
}

\abstract{We study boundary scattering in the $\phi^4$ model on a
half-line with a one-parameter family of Neumann-type boundary
conditions. A rich variety of phenomena is observed, which extends
previously-studied behaviour on the full line to include regimes of
near-elastic scattering, the restoration of a missing scattering
window, and the creation of a kink or oscillon through the
collision-induced decay of a metastable boundary state.
We also study the decay of the vibrational boundary mode, and explore
different scenarios for its relaxation and for the creation of kinks.
}


\maketitle

\section{Introduction}
%
Systems with boundaries, defects and impurities have been
intensively studied
in statistical physics and field theory,
both at the classical and the quantum levels.
Often the key physics of
the model can be captured, possibly after dimensional reduction,
 by a simple 1+1 dimensional field
theory on a half line.  Examples include the
Kondo problem \cite{Kondo},
fluxon propagation in long Josephson junctions \cite{Olsen:1981a},
the XXZ model with boundary magnetic field
\cite{Alcaraz:1987},
an impurity in an interacting electron gas
\cite{Kane:1992zza},
the sine-Gordon \cite{Ghoshal:1993tm} and Toda \cite{Bowcock:1995vp} models,
monopole catalysis \cite{Rubakov:1981rg},
the Luttinger liquid \cite{Wen:1990se},
and a toy model motivated by M-theory
\cite{Antunes:2003kh}.

Especially since the work of Ghoshal and
Zamolodchikov \cite{Ghoshal:1993tm}, there has been great interest
in boundary conditions compatible with
bulk integrability, and many such models turn out to
be of direct physical interest. However less attention has been paid
to the equally if not more physically-relevant cases of non-integrable
boundary systems, even at the classical level. This is perhaps a
shame, as it is now known that non-integrable classical field
theories, even in 1+1 dimensions, can exhibit remarkably rich
patterns of behaviour not seen in their integrable counterparts
\cite{Peyrard:1984qn,Campbell:1983xu,Anninos:1991un,Goodman:2007,
Dorey:2011yw}.

In this paper we examine the $\phi^4$ theory in 1+1 dimensions,
restricted to a half line by a simple Neumann-type `magnetic field'
boundary condition. (The sine-Gordon model with a non-integrable
boundary was recently investigated in 
\cite{Arthur:2015mva}.)
The $\phi^4$ theory on a full line
is similar to the sine-Gordon model in that both
support topological kinks and antikinks;
the $\phi^4$ theory also has an intriguing and still not fully-understood
counterpart of the sine-Gordon breather,
the oscillon \cite{Bogolyubsky:1976nx}. We chose the magnetic
field boundary condition in part because of its simplicity, and in
part
because the scattering of kinks against such a boundary provides a natural
deformation of the full-line scattering problems which are already
known to
exhibit intricate patterns of resonant scattering
\cite{Peyrard:1984qn,Campbell:1983xu,Anninos:1991un,Goodman:2007}.
In some regimes our results do indeed
resemble
the pattern of scattering windows observed in
kink-antikink collisions on the full line, while
in others we find novel phenomena including
a new type of `sharp-edged' scattering window. Even though
the theory is not
integrable, it turns out to be possible
to give an accurate analytical description of some aspects of
this behaviour. We complement these studies with an investigation of
the decay of the vibrational boundary mode through nonlinear couplings
to scattering states, and of the creation of kinks by an excited
boundary.  An interesting feature of the boundary mode decay,
discussed in section~\ref{sec:higherorder}, is that with
suitable initial conditions a period of relatively slow decay
can be followed by a sudden burst of radiation from the boundary as a
new decay channel opens.

While this paper is self-contained, we have also made a number of
short movies to illustrate aspects of the discussion, which can be found
at \cite{animations}.  After a brief explanation of the numerical
methods used to obtain our plots in appendix \ref{app:methods}, these
are listed in appendix~\ref{app:supplementary}.

\smallskip
\section{The model}
We consider a rescaled $\phi^4$ theory with vacua
$\phi_v \in \{-1,+1\}$ on the left half-line $-\infty < x < 0$. The
bulk energy and Lagrangian densities are
$\mathcal{E}=\mathcal{T}+\mathcal{V}$ and
$\mathcal{L}=\mathcal{T}-\mathcal{V}$ respectively, where
\begin{equation}
  \label{Lag}
\mathcal{T}=\halft \phi_t^2~~\mbox{and}~~
\mathcal{V}=\halft \phi_x^2+\halft(\phi^2-1)^2\,.
\end{equation}
The static full-line kink and antikink,
$\phi_K(x) = \tanh (x-x_0)$ and $\phi_{\bar K}(x)=-\phi_K(x)$,
have rest mass $M=4/3$ and interpolate between the two vacua.
Including a boundary energy $-H\phi_0$, where $\phi_0=\phi(0,t)$ and
$H$ can be interpreted as a boundary magnetic field,
yields the Neumann-type boundary
condition $\phi_x(0,t)=H$ at $x=0$\,.

\begin{figure}
\centering
 \hspace*{-1pt}\includegraphics[width=0.8\textwidth,angle=0]{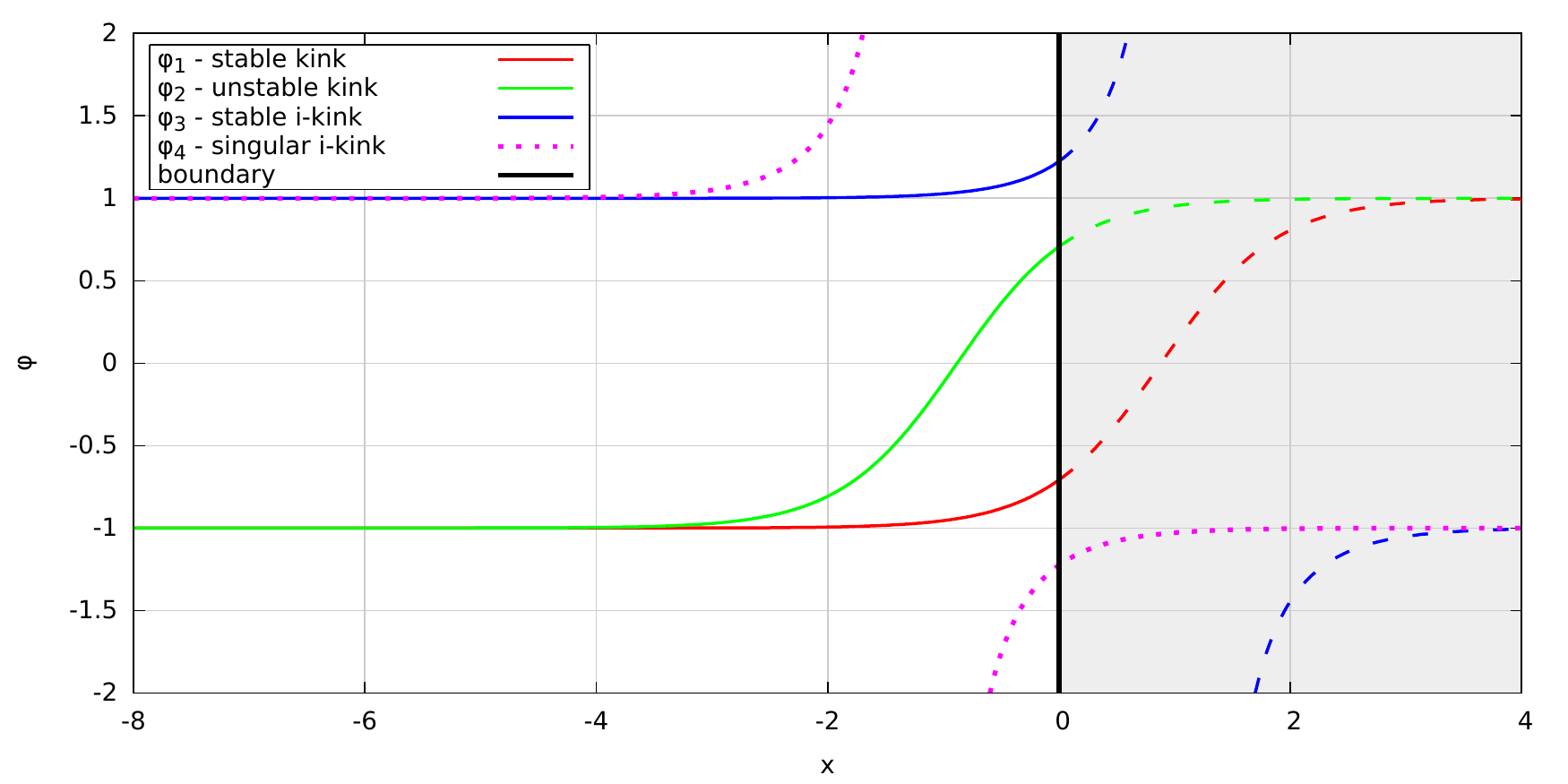}
 \vspace{-10pt}
 \caption{\label{fig:static}\small Static solutions for $H=1/2$.}
 \vspace{-10pt}
\end{figure}

For $0<H<1$ there are four static
solutions
to the equations of motion,
shown in figure \ref{fig:static}. Two of them,
$\phi_1(x)=\tanh(x-X_0)$ and $\phi_2(x)=\tanh(x+X_0)$
with $X_0=\cosh^{-1}(1/\sqrt{|H|})$, are restrictions of regular full-line
kinks to the half-line,
while the other two, $\phi_3(x)=-\coth(x-X_1)$ and
$\phi_4(x)=-\coth(x+X_1)$ with $X_1=\sinh^{-1}(1/\sqrt{|H|})$ are
irregular on the full line. On the
half line, $\phi_3$ is non-singular and corresponds to the
absolute minimum of the energy, while
$\phi_1$ is metastable, and $\phi_2$ is the unstable
saddle-point between $\phi_3$ and $\phi_1$.
Their energies
can be found by rewriting $E[\phi]
=\int_{-\infty}^0\mathcal{V}\,dx-H\phi_0$ in Bogomolnyi form as
\begin{equation}
E[\phi]=\fract{1}{2}\!\!\int\limits_{-\infty}^0 \!\!
\left(\phi_x\pm(\phi^2{-}1)\right)^2\! dx
\mp\left[\fract{1}{3}\phi^3{-}\phi\right]_{-\infty}^0
 - H\phi_0\,.
\label{BPS}
\end{equation}
Since $\phi_1$ and $\phi_2$ satisfy
$\phi_x=1-\phi^2$ we have
$\phi_1(0)=-\sqrt{1{-}H}$, $\phi_2(0)=\sqrt{1{-}H}$\,;
while
$(\phi_3)_x=\phi_3^2-1$ and so $\phi_3(0)=\sqrt{1{+}H}$.
Taking the upper and lower signs in (\ref{BPS}) as appropriate,
\begin{align}
E[\phi_1]&= \fract{2}{3}-\fract{2}{3}(1{-}H)^{3/2}\,,~~~
E[\phi_2]= \fract{2}{3}+\fract{2}{3}(1{-}H)^{3/2}\,, \nonumber\\
E[\phi_3]&= \fract{2}{3}-\fract{2}{3}(1{+}H)^{3/2}\,.
\end{align}

As $H$ increases through $1$,
$\phi_1$ merges with $\phi_2$ and disappears, leaving $\phi_3$ as the
only static solution for $H>1$. For $H<0$ the story is the same,
with $\phi$ and $H$ negated throughout, so the physically-relevant
solutions are $\tilde\phi_i(x):=-\phi_i(x)$, $i=1\dots 3$.

\smallskip

\section{Numerical results}
We took initial conditions corresponding to an
antikink at $x_0=-10$
travelling towards the boundary with velocity $v_i>0$.
(We found the setup with an incident antikink easier to
visualise, but our results apply equally to kink-boundary collisions
on negating $\phi$ and $H$.) Thus
the initial profile
was $\phi(x,0)=\phi_1(x)-\tanh(\gamma(x-x_0))+1$
for $H>0$ and
$\phi(x,0)=\tilde\phi_3(x)-\tanh(\gamma(x-x_0))+1$
for $H<0$, where $\gamma=1/\sqrt{1-v_i^2}$.
Our real interest was in the
problem with the initial antikink
infinitely far from the boundary, but the rapid decay of the
antikink-boundary force (\ref{force}), calculated below,
meant that error in taking $x_0=-10$ was small.

To solve the system numerically, we restricted it to an
interval of length $L$, with the Neumann boundary condition imposed
at $x=0$ and a Dirichlet condition at $x=-L$.
(Since we took run times such that radiation did not have time to reflect from
the extra boundary and return, the
boundary condition at $x=-L$ was anyway irrelevant.)
We used
a 4\textsuperscript{th} order finite-difference method, explained in
more detail in appendix \ref{app:methods},
on a grid of $N=1024$ nodes with $L=100$, so the spatial step was
$\delta x\approx 0.1$, and a
6\textsuperscript{th}-order
symplectic integrator
for the time stepping function, with
time step $\delta t = 0.04$. Selected runs were repeated with
other values of $x_0$, $L$, $N$ and $\delta t$
to check the stability of our results.

\begin{figure}
 \centering
 \hspace*{-1pt}\includegraphics[width=1\textwidth,angle=0]{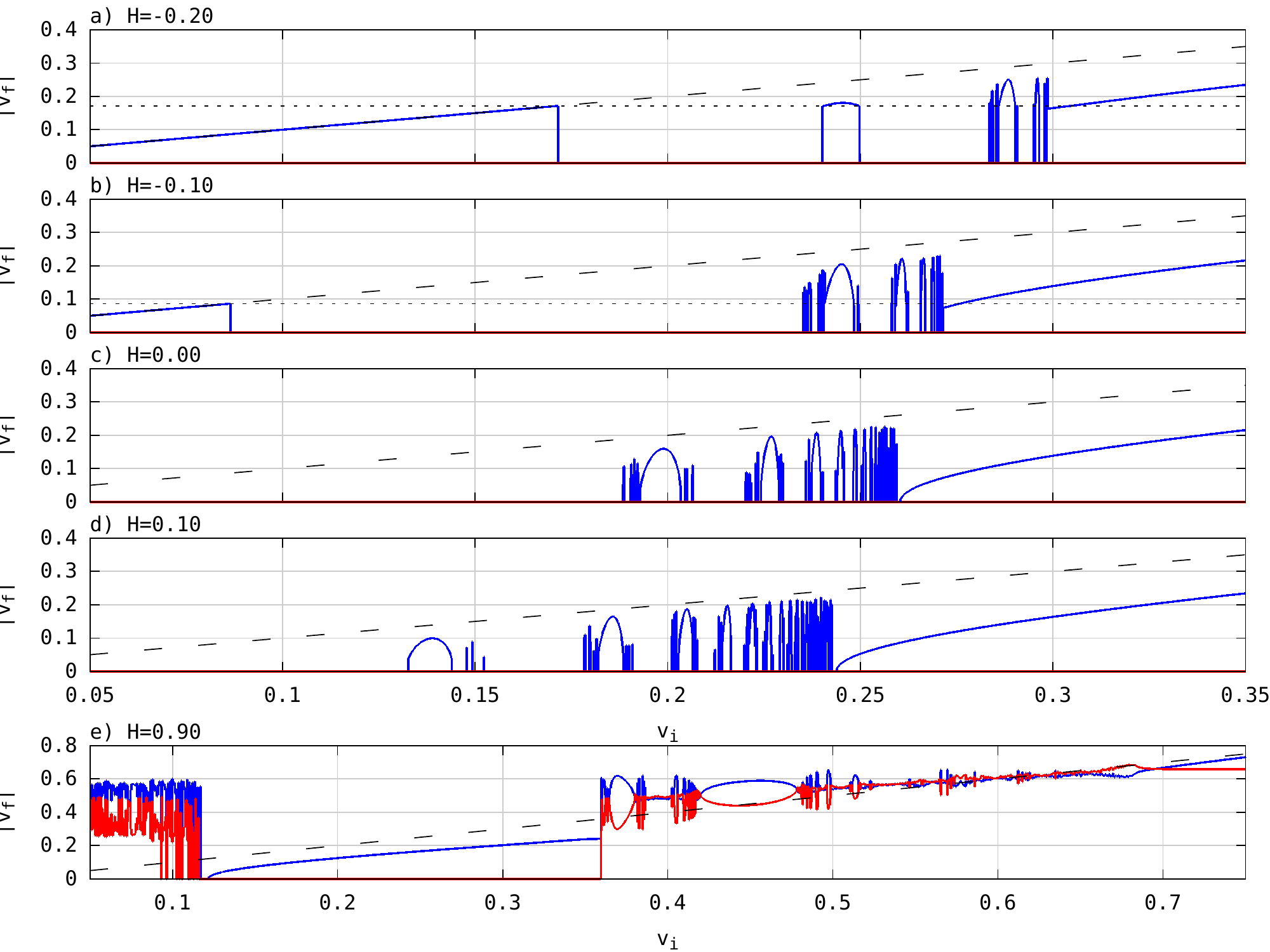}
 \vspace{-14pt}
 \caption{\label{fig:kb}\small
Final antikink velocities as functions of initial
velocities. The dashed line indicates the result for a purely
elastic collision. In the fifth plot, a kink can also be
produced: its velocity is shown in red. The horizontal dotted lines in
plots a and b show the relevant
values of $v_{cr}(H)$, as given by equation
(\ref{eq:vcr}) below.
}
 \vspace{1pt}
\end{figure}
\begin{figure}
 \begin{center}
  \hspace*{-1pt} \includegraphics[width=1\textwidth,angle=0]{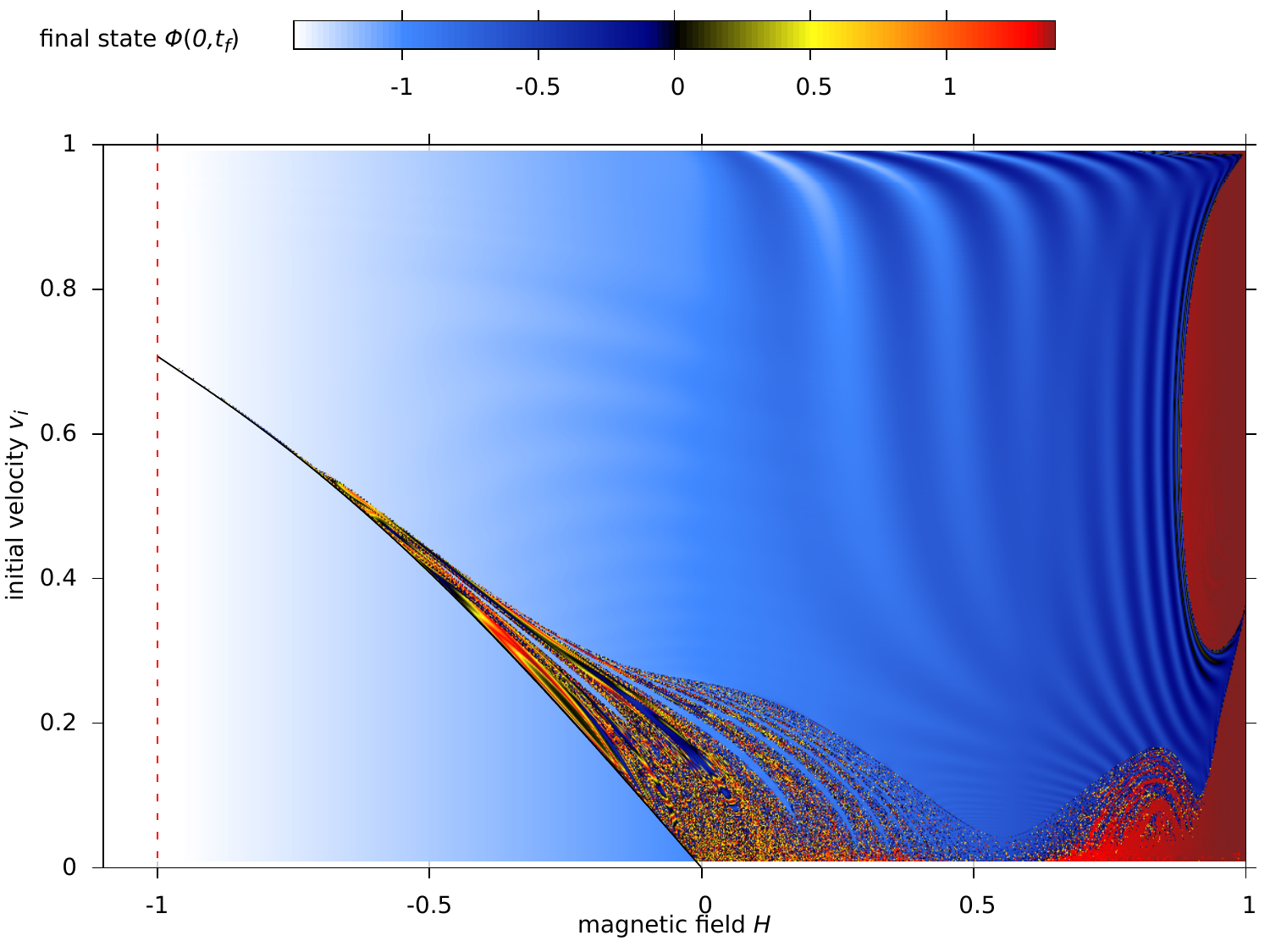}
 \vspace{-8pt}
 \caption{\label{fig:scan}\small A `phase diagram' of
antikink-boundary collisions. The plot
shows the value of the field at $x=0$ a time
$t_f=|x_0|/v_{i}+100$ after the start of the simulation,
as function of the boundary magnetic field $H$ and the
initial velocity $v_{i}$.}
\end{center}
 \vspace{-7pt}
\end{figure}

Our simulations revealed a rich picture, aspects of
which are summarised in figures~\ref{fig:kb} and~\ref{fig:scan}. For all
$(H,v_i)$ pairs with
$H<H_c\approx 0.6$, the antikink either
reflects off the boundary with some velocity $v_f$, or becomes stuck
to it -- corresponding to $v_f=0$ -- to form a `boundary oscillon'.
This latter configuration oscillates with a large (of order one)
amplitude, and a below-bulk-threshold basic frequency.
Just like the bulk oscillon (which it becomes in the
limit $H\to 0$), it then decays very slowly into radiation.
At $H=0$ (figure~\ref{fig:kb}c) the plot of $|v_f|$ as a function of $v_i$
reproduces the well-known structure of resonant scattering windows
in $K\bar K$ collisions on a full line
\cite{Peyrard:1984qn,Campbell:1983xu,Anninos:1991un}.
For \textit{negative} values of $H$ (figures \ref{fig:kb}a and b)
new features emerge.
For $v_i$ small, the antikink is reflected elastically
from the boundary with very little radiation.
As $v_{i}$ increases above an $H$-dependent
critical value $v_{cr}$, the antikink is trapped
by the boundary, leaving only radiation in the final state. Increasing $v_{i}$
further, scattering windows begin to open, until $v_i$ exceeds
an upper critical value and the antikink again always escapes.
If the antikink does escape, its speed
$|v_f|$ is always larger than some minimal value
very slightly lower than $v_{cr}$, so (in contrast to the
full-line situation)  $v_f$ is a discontinuous function of $v_i$,
giving the windows the sharp edges mentioned in the introduction.
For small \textit{positive} values of $H$ (figure \ref{fig:kb}d),
$v_f$ is instead a continuous function of $v_i$, 
the sequence of windows for $H=0$ shifting towards
lower velocities while preserving its general
structure.
Finally, for $H>H_c$
(figure \ref{fig:kb}e)
other new phenomena arise which have no counterparts in the full-line
theory; these will be discussed further in later sections.

\smallskip

\section{Kink-boundary forces and the location of the low-velocity
window}
To understand the novel window of near-elastic scattering 
at low initial velocities when $H$ is negative,
seen in figures \ref{fig:kb}a and
\ref{fig:kb}b, we start by
evaluating the static force between a single antikink and the
boundary.  Placing the antikink at $x=x_0<0$, we add a
possibly-singular `image'
kink at $x_1>0$ in such a way that the combined configuration satisfies
the boundary condition at $x=0$. From the standard full-line result,
the force on the antikink from the image kink
is equal to $32e^{-2(x_1-x_0)}$, or minus
this if the image kink is singular.
For $|H|\ll 1$ and $|x_0|\gg 1$ we find that the boundary condition
requires
$e^{-2x_1}=\frac{1}{4}H+e^{2x_0}$, so
\begin{equation}
F = 32 \left( \fract{1}{4} H +e^{2x_0} \right)e^{2x_0}.
\label{force}
\end{equation}
For $H<0$ the
force is repulsive far from the boundary,
only becoming attractive nearer in.
When $x_0=\frac{1}{2}\log(-\frac{1}{4}H)$, $x_1=\infty$ and the force
vanishes,
the antikink-kink
configuration
reducing to the unstable static solution $\tilde\phi_2$.

\begin{figure}
 \begin{center}
\hspace*{-5pt}
  \includegraphics[width=1.0\textwidth,angle=0]{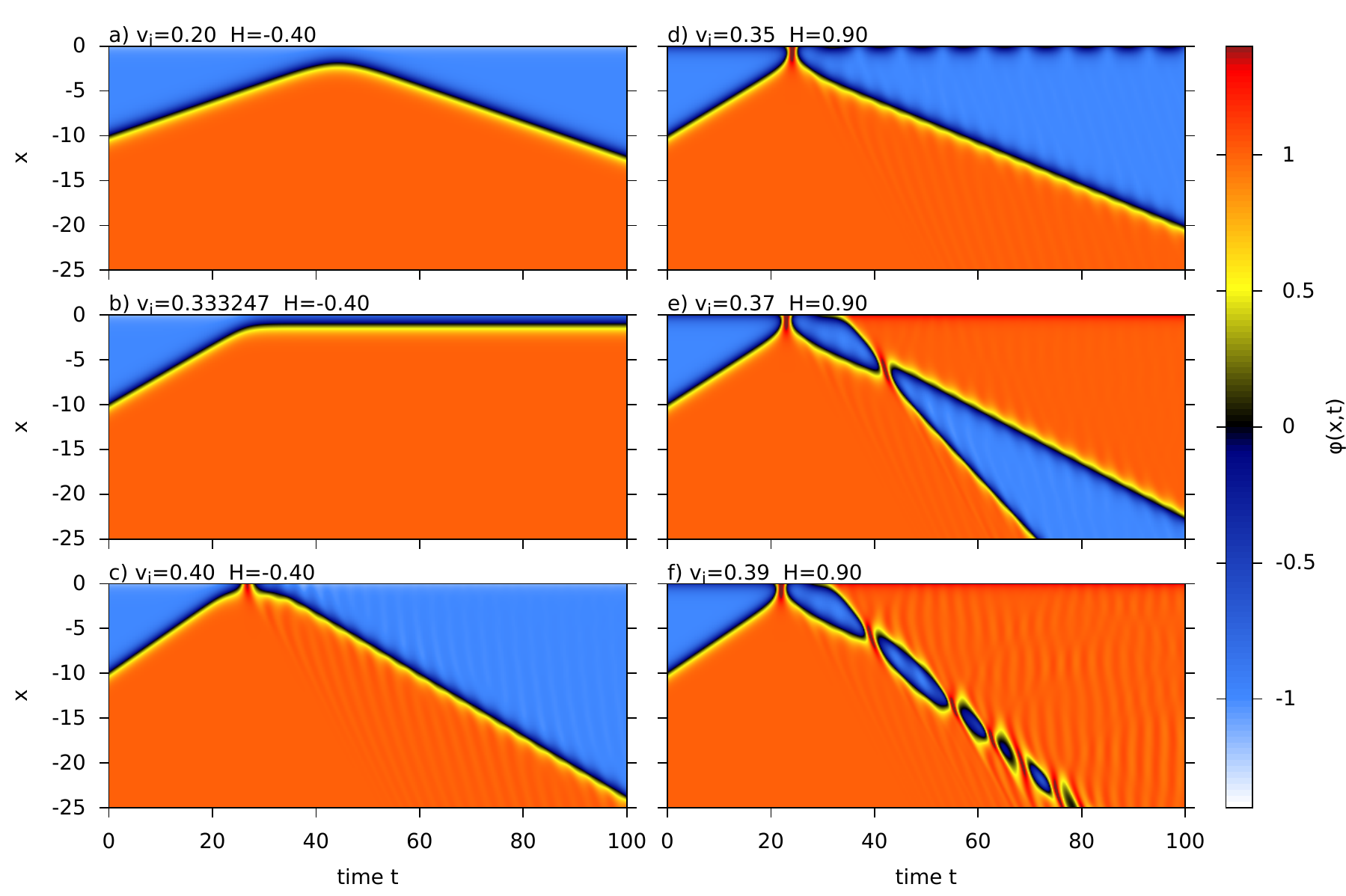}
 \vspace{-9pt}
 \caption{\label{fig:examples}\small
Example collisions for $H=-0.4$ (left column) and $H=0.9$ (right
column),
illustrating various scattering scenarios. At negative $H$:
(a) elastic recoil for low impact velocity;
(b) saddle point production at the critical velocity, the antikink 
finishing on the top of the
potential barrier; (c) single bounce with the antikink escaping back
over the barrier. At positive $H$:  (d) single
bounce with the excitation of the $H>0$ boundary mode;
(e) kink production via collision-induced boundary decay;
(f) bulk oscillon production. See also movies
\texttt{M01} -- \texttt{M07} and \texttt{M11}
of appendix \ref{app:supplementary}.}
\end{center}
 \vspace{-10pt}
\end{figure}

\begin{figure}[t]
\hskip -15pt
\includegraphics[width=1.05\textwidth]{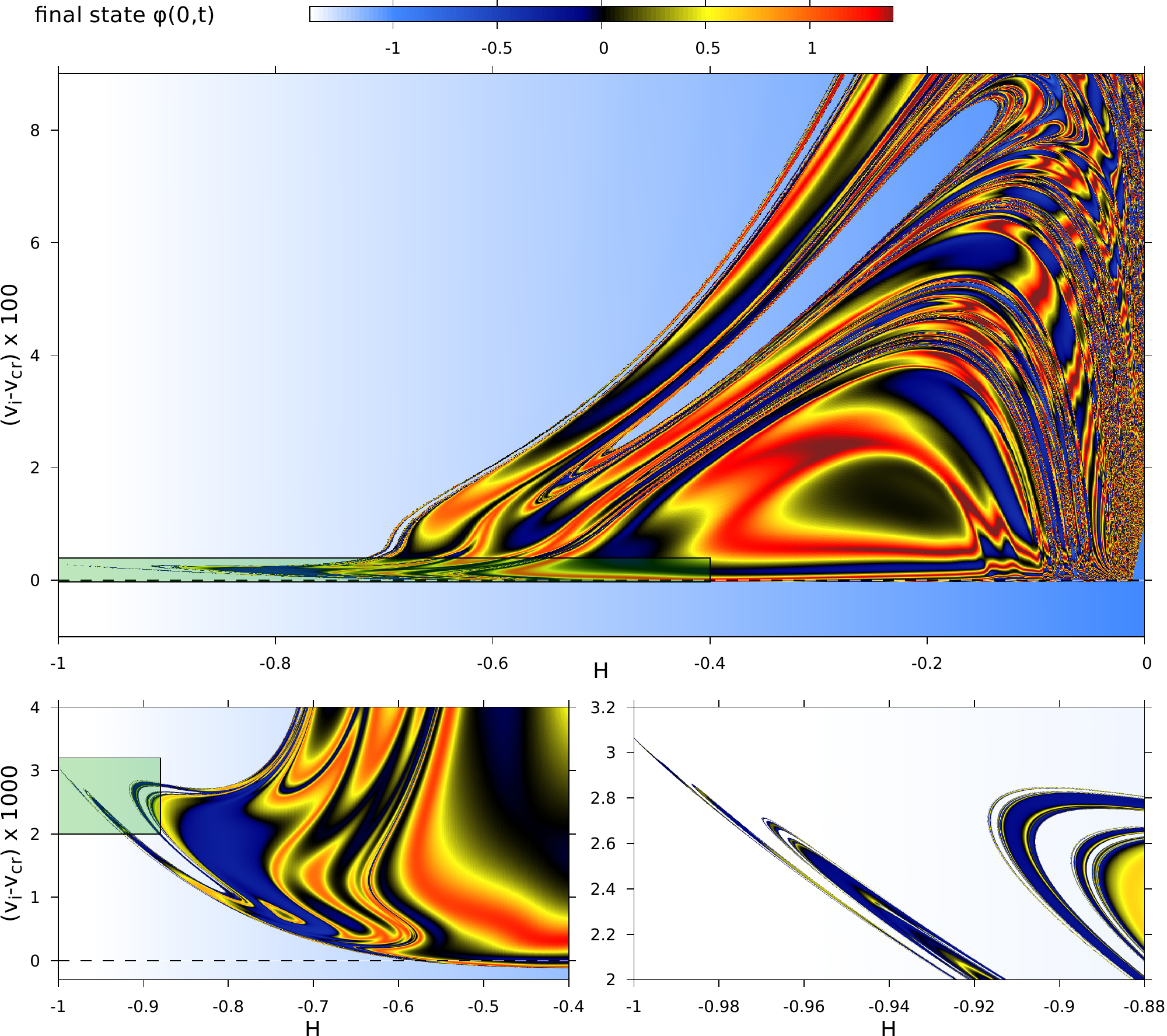}
 \vspace{-25pt}
 \caption{\label{fig:saturn}\small Zoomed-in views of the region near
the tip of the fractal tongue of figure \ref{fig:scan}. Note that the
vertical axes show multiples of $v_i-v_{cr}$, where $v_{cr}=v_{cr}(H)$ is 
the theoretical upper limit of the near-elastic scattering window given
by the formula (\ref{eq:vcr}).}
\end{figure}

Now consider, again for $H<0$, an antikink moving towards
the boundary. If its velocity $v_i$ is small, then it won't have
sufficient energy to overcome the initially-repulsive force,
and it will be reflected without ever coming close to
$x=0$, and without significantly exciting any other modes; this
behaviour
is illustrated in figure \ref{fig:examples}a. Increasing $v_i$, at some
critical value $v_{cr}$ the energy will be just enough reach the top
of the potential barrier and create the static saddle-point configuration
$\tilde\phi_2$, as shown in figure \ref{fig:examples}b.
The value of $v_{cr}$ can be deduced on energetic
grounds: the initial energy is
$\frac{4}{3}(1-v_{cr}^2)^{-1/2}+E[\tilde\phi_3]$, while the final energy is
$E[\tilde\phi_2]=\frac{2}{3}+\frac{2}{3}(1{+}H)^{3/2}$. Equating the
two,
\begin{equation}
v_{cr}(H)=\sqrt{1-4 \bigl((1{+}H)^{3/2}+(1{-}H)^{3/2}\bigr)^{-2}}\,.
\label{eq:vcr}
\end{equation}
If $v_i$ is just larger than $v_{cr}$, the antikink can
overcome the potential barrier and approach the
boundary; energy is then lost to other modes and so it is
unable to return, and is trapped at the boundary. Thus
$v_{cr}(H)$ marks the upper limit of the windows of
almost-perfectly-elastic scattering seen in
figures \ref{fig:kb}a and \ref{fig:kb}b, and the lower edge of the `fractal
tongue' occupying the left half of figure \ref{fig:scan}.
The curve $v_i=v_{cr}(H)$ is included in figure \ref{fig:scan};
it matches our numerical results remarkably well. 
Indeed, it can be seen from figure \ref{fig:saturn} below that the 
maximum error is of the
order of $0.5\%$, which is rather 
small given that radiation was ignored in the derivation.
Similar arguments show that,
within this approximation, $v_{cr}$ is the smallest
possible speed for \textit{any} escaping antikink, explaining
the sharp (discontinuous) edges of all windows when $H<0$.

\section{The boundary mode}
\begin{figure}
\hspace* {-15pt}
\includegraphics[width=1.03\textwidth,angle=0]{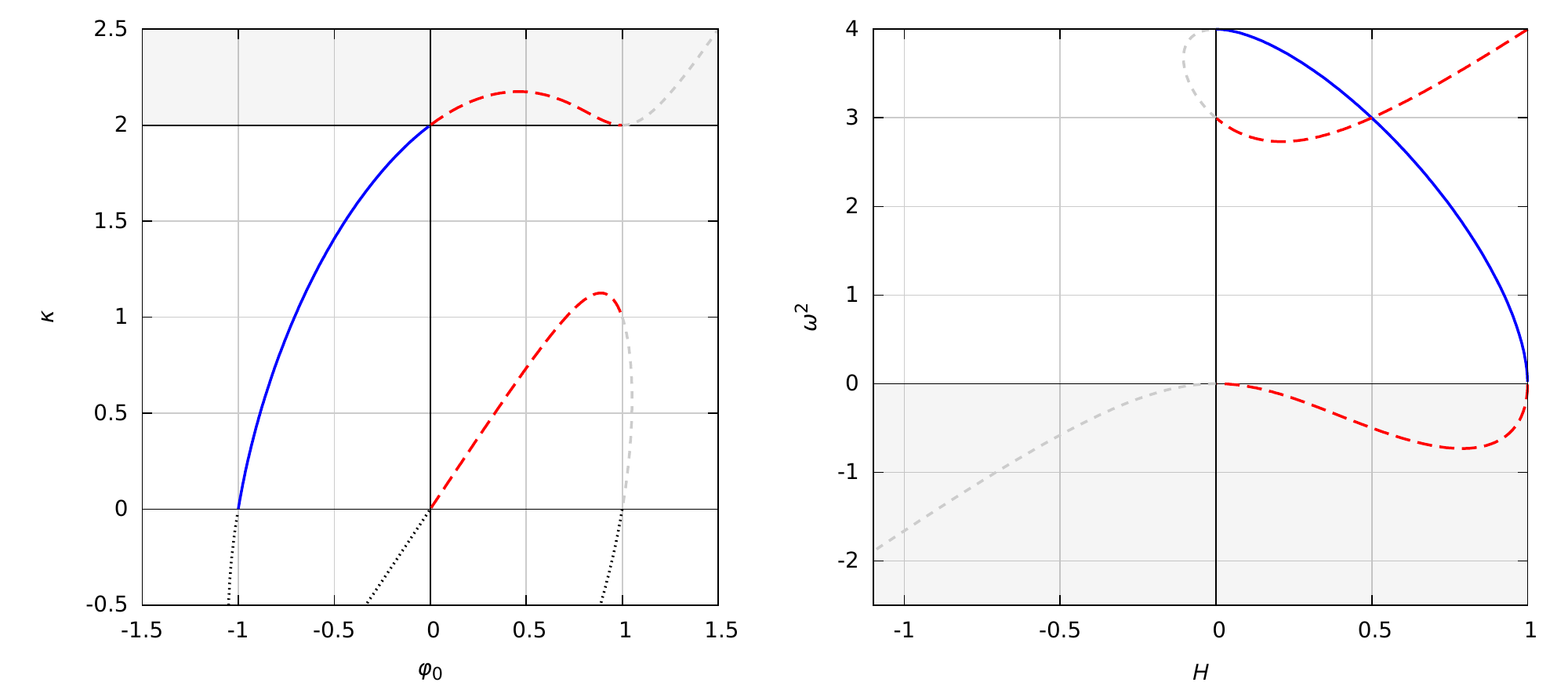}
\caption{\label{fig:modes}\small
Linearised boundary mode analysis: on the left, the solutions of equation
(\ref{eq:kappa}) as a function of $\phi_0=\phi(0)$; on the right, the
frequencies of localised boundary modes as a function of $H$.}
\end{figure}

Next we consider the perturbative sector of the model, that is
solutions of the form $\phi(x,t)=\phi_s(x)+\eta(x,t)$ where
$\phi_s(x)$ is a static solution to the equations of motion
and $\eta(x,t)$ is small.
The full-line theory has a continuum of small linear
perturbations about each vacuum with
mass $m = 2$, while
a static kink $\phi_K(x)=\tanh(x-X_0)$ has two discrete normalizable
modes -- the translational mode, and a vibrational mode
with frequency $\omega_1 = \sqrt 3$ -- and a continuum of
above-threshold states $\eta(x,t)=e^{i\omega t}\eta_k(x)$
where $\omega^2=4+k^2$ and \cite{Sugiyama:1979mi}
\begin{equation}
\eta_k(x) = e^{-ik (x-X_0)}\left(-1-k^2+3ik\tanh (x-X_0) + 3\tanh^2
(x-X_0)\right).
\label{pertfun}
\end{equation}
Turning now to the half-line theory, we can
regard $\phi_K(x)$ instead as the static half-line solution $\phi_1(x)$
to the boundary theory with $0<H<1$ and $\phi(-\infty)=-1$. 
Its linear perturbations 
must now satisfy $\partial_x \eta(x)=0$ at $x=0$.
Setting $k=i\kappa$ this yields
\begin{equation}
\kappa^3-3\phi_0\kappa^2+(6\phi_0^2-4)\kappa-6\phi_0^3+6\phi_0=0
\label{eq:kappa}
\end{equation}
where $\phi_0=\phi_1(0)=-\sqrt{1-H}$ and the frequency $\omega_B$ of the
corresponding boundary mode satisfies $\omega_B^2=4-\kappa^2$. 
The solutions of (\ref{eq:kappa}) for both negative and positive
values of $\phi_0$ are shown on the left-hand plot
of figure \ref{fig:modes};
note that only solutions with $\kappa>0$ can give rise to localised
modes, and of these, $\kappa$ must be less than $2$ for
$\omega_B$ to be real and the mode stable.
We will denote the corresponding normalised profile function as
$\eta_B(x):=\eta_{i\kappa}(x)/\eta_{i\kappa}(0)$, where
$\eta_{i\kappa}(x)$ is given by (\ref{pertfun}) with $k=i\kappa$.

For $0<H<1$,
we have $-1<\phi_0<0$
and (\ref{eq:kappa}) has just
one positive solution $\kappa$, which 
satisfies $\kappa<2$: this is the single vibrational mode,
localised near to the boundary.
The linear perturbations of $\phi_2(x)$, the saddle-point solution,
are also described by (\ref{eq:kappa}), but now with
$\phi_0=\phi_2(0)=+\sqrt{1-H}$.
 For these
cases (\ref{eq:kappa}) has two positive solutions but one
is larger than $2$: this is the unstable mode of $\phi_2(x)$.
Finally, for $H<0$,
the continuation of (\ref{eq:kappa})
to $\phi_0<-1$
governs the spectrum of fluctuations about $\tilde\phi_3(x)$, the
$H<0$ vacuum in the $\phi(-\infty)=-1$ sector.
There are no positive solutions in this regime and hence no
vibrational
modes of the boundary for $H<0$. The right-hand plot
of figure~\ref{fig:modes} summarises the situation, plotting the
images of the positive-$\kappa$ parts of the curves shown on the left under
the mapping $(\phi_0,\kappa)\to (H,\omega_B^2)=(1{-}\phi_0^2,4{-}\kappa^2)$.
The grey dashed parts of the curve visible for $H<0$ are included for
completeness but do not describe vibrational modes of physical
solutions -- they correspond to `perturbations' of
the singular solution $\tilde\phi_4(x)$.

\begin{figure}
 \begin{center}
   \includegraphics[width=0.8\textwidth,angle=0]{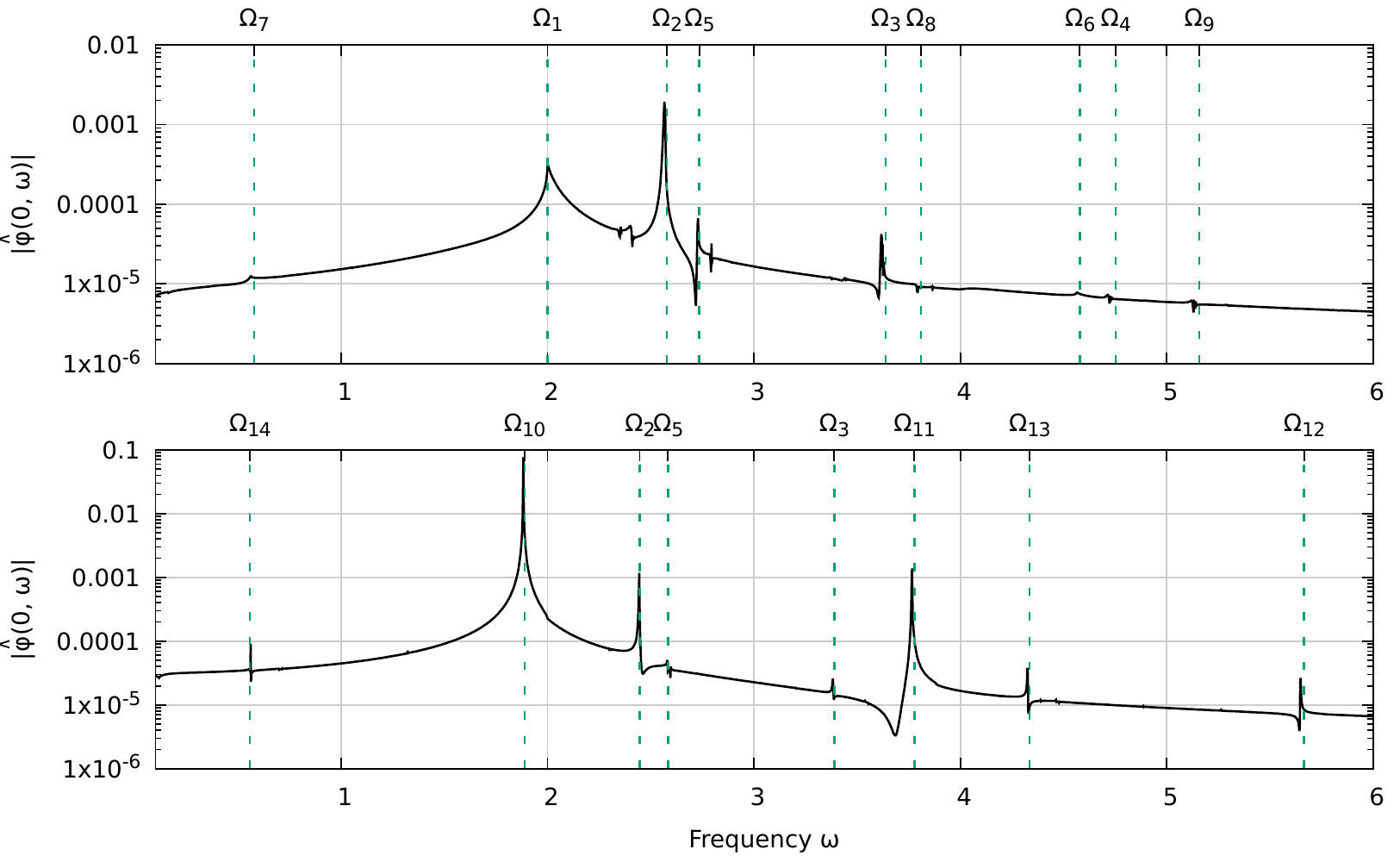}
 \vspace{-7pt}
 \caption{\label{fig:spectra}\small Power spectra at the boundary
after a collision with $v_{i}=0.5$, for $H=-0.1$ (upper)
 and $H=0.3$ (lower).}
\end{center}
 \vspace{-14pt}
\end{figure}

These findings are confirmed by our numerical results.
Figure \ref{fig:spectra} 
shows the Fourier transforms of $\phi(0,t)$ for $30<t<3030$,
for antikink-boundary
collisions with initial velocity $v_i=0.5$, and
$H=-0.1$ and $0.3$.
The final velocity $v_f$
of the reflected antikink is $-0.382596$ for
$H=-0.1$ and $-0.454014$ for $H=0.3$, so in both cases translational
energy is lost to other modes during the collision.

For $H=-0.1$, the boundary does not have an internal mode, and
only radiative modes with frequencies near to $2$, the mass threshold,
remain near to the boundary.
The internal mode of the reflected antikink has
frequency $\omega_1$, but this mode cannot be
observed at the boundary since it is exponentially suppressed there.
However nonlinear couplings with other excitations create waves with
frequencies at above-threshold multiples of $\omega_1$
\cite{Manton:1996ex}, which can propagate back to the boundary.
The upper plot of figure \ref{fig:spectra} shows peaks
at $\Omega_1=2$ and $\Omega_2=\Omega(2\omega_1)$, where
$\Omega(\omega)=\gamma(\omega+k(\omega)v_f)$ is the Doppler-shifted
frequency of radiation emitted from the moving kink measured on
the boundary. Higher harmonics at $\Omega_3=\Omega(3\omega_1)$ and
$\Omega_4=\Omega(4\omega_1)$ are also visible, along with combinations of
the internal mode of the antikink and the lowest continuum mode such
as $\Omega_5 = \Omega(2+\omega_1)$ and $\Omega_6=2+\Omega(2\omega_1)$.

Many of these modes are also present in the $H=0.3$ spectrum shown in
the lower plot of figure \ref{fig:spectra}, albeit at shifted locations
because of the different final antikink velocity.
However the plot is dominated by the internal boundary mode with
frequency $\Omega_{10}=\omega_B=1.888459$. The
higher harmonics $\Omega_{11}=2\omega_B$ and
$\Omega_{12}=3\omega_B$ are also visible, while
interactions between radiation from
the outgoing antikink and the boundary mode lead to
peaks at $\Omega_{13} =\omega_B+\Omega(2\omega_1)$ and $\Omega_{14}
=\omega_B-\Omega(2\omega_1)$.

\smallskip

\section{The resonance mechanism in boundary scattering}
For small nonzero values of $|H|$, the resonant energy
exchange mechanism governing scattering in the bulk $\phi^4$ model is
changed in two ways in the boundary theory:
(i) the attractive force acting on the antikink
near to the boundary is modified, in particular becoming repulsive at
greater distances when $H$ is negative; (ii)
after the initial impact, energy can be stored not only in the internal
mode of the antikink, but also, for positive values of $H$,
in the boundary mode. These factors change the resonance condition
for energy to be returned to the translational mode of the
antikink on a subsequent impact after some integer number of
oscillations of the antikink's internal mode, shifting (and,
for negative $H$, sharpening) the windows seen in
figures \ref{fig:kb}\,a-d. This return can happen after two, three or
more bounces from the boundary, leading to a hierarchy of multibounce
windows as in the full-line situation.
Our numerical results suggest that
for small positive values of $H$ the contribution of the
boundary mode in the resonant energy transfer is not significant.

\begin{figure}
 \begin{center}
 \hspace*{-19pt}
   \includegraphics[width=1.07\textwidth,angle=0]{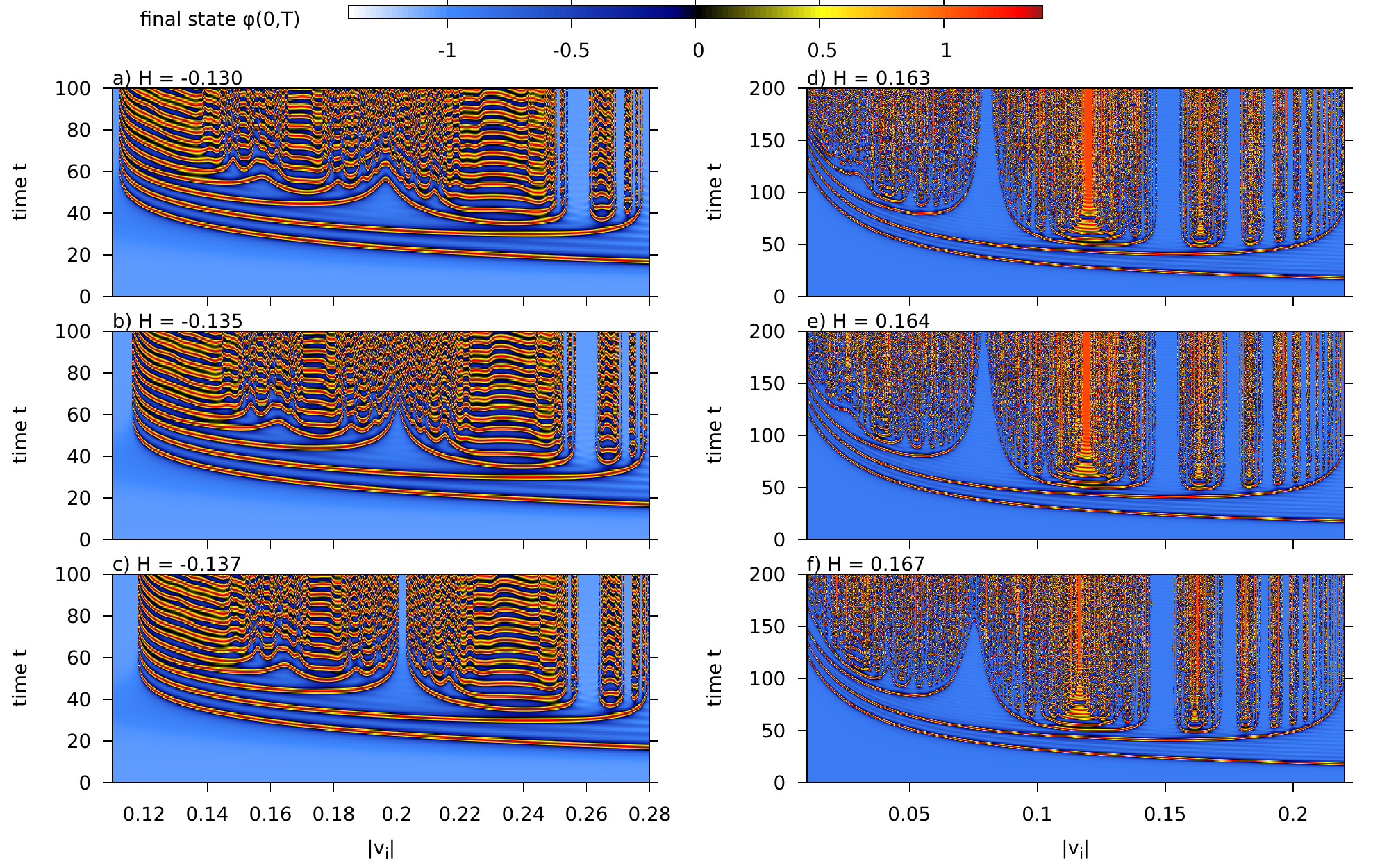}
 \vspace{-19pt}
 \caption{\label{fig:windowResurrections}\small 
Scans of $\phi(0,t)$ for various values of $H$, showing the
emergence (on the left) and destruction (on the right) of a 
two-bounce window as $H$ moves away from zero.}
\end{center}
 \vspace{-10pt}
\end{figure}

\begin{figure}
 \begin{center}
   \includegraphics[width=0.87\textwidth,angle=0]{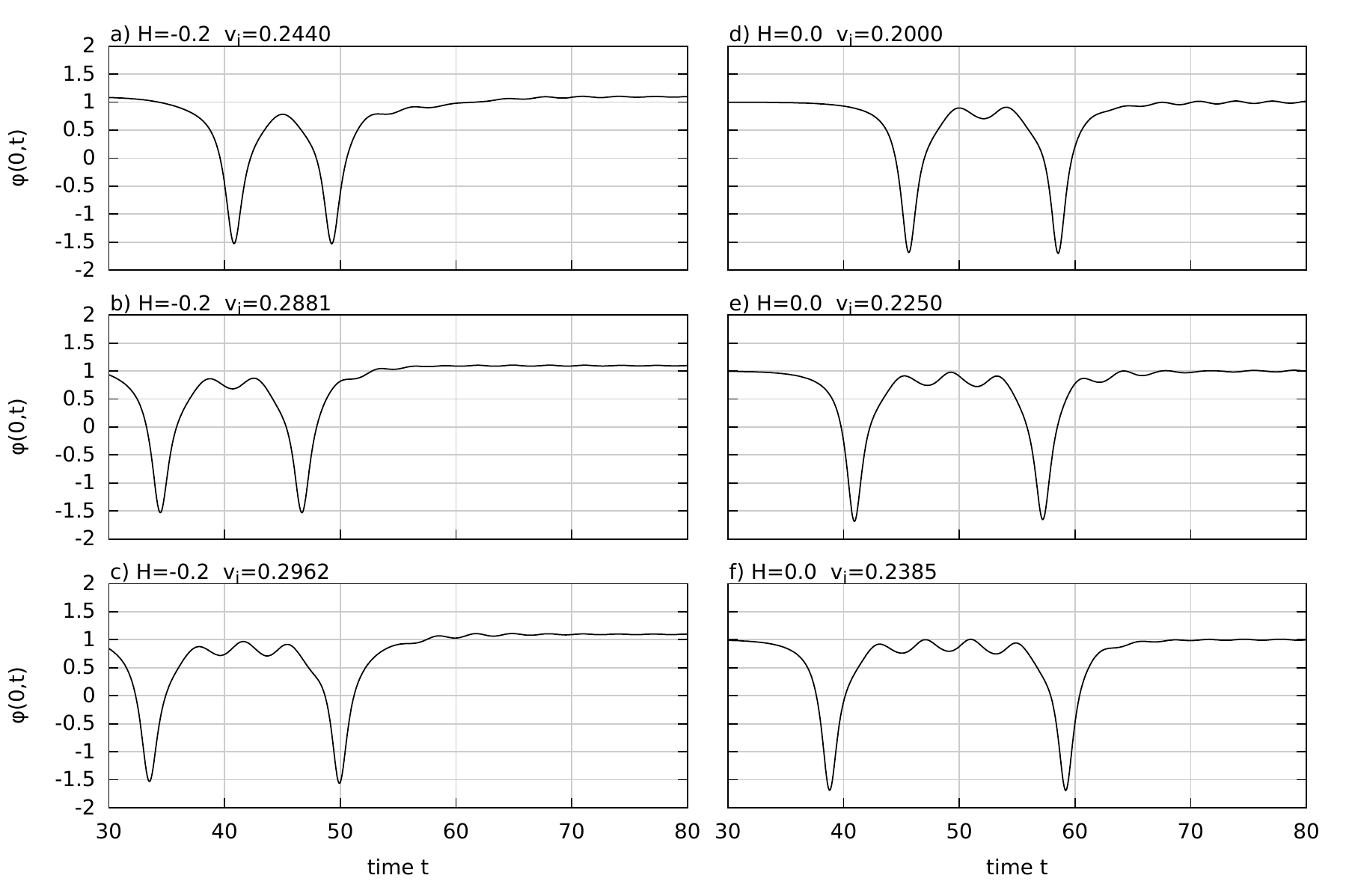}
 \vspace{-5pt}
 \caption{\label{fig:bounces}\small Plots of $\phi(0,t)$ for $v_i$
inside the first three two-bounce windows for $H=-0.2$
(left) and for $H=0$ (right).}
\end{center}
 \vspace{-15pt}
\end{figure}

For larger values of $|H|$ other new features appear.
For $H<0$ the first is the 
resurrection of a two-bounce window that was observed to
be missing from the
full-line scattering process by Campbell et al in \cite{Campbell:1983xu}.
Figure \ref{fig:kb}a
includes a scattering window centred at $v_i\approx 0.245$ which is
\textit{not}
the continuation of any of the windows seen in
figures \ref{fig:kb}\,b-d; the same window can be seen in
figure \ref{fig:scan} running from
$(H,v_i)=(-0.135,0.202)$ to $(H,v_i)=(-0.489,0.417)$, 
and the top plot of figure \ref{fig:saturn}, running from
$(H,v_i-v_{cr})=(-0.135,0.084)$ to $(H,v_i)=(-0.489,0.023)$.
The emergence of this window as $H$ decreases below $H\approx -0.136$
is shown in more detail in the left-hand set of plots of
figure \ref{fig:windowResurrections}.
The nature of the new window is made clear by the plots in
figure~\ref{fig:bounces}, which
shows $\phi(0,t)$ for $v_i$ inside the first three two-bounce
windows for $H=-0.2$, and also for $H=0$, which is equivalent to the 
full-line case. 
The `wobbles' between the large dips in such
plots count the oscillations of the internal antikink mode between
bounces \cite{Campbell:1983xu}.
As can be seen from the figure, the minimum number of
oscillations supporting antikink escape is one smaller for
$H=-0.2$ than it was for $H=0$, giving rise to the extra window. 
A complementary process of `window destruction' occuring for $H>0$ can
be seen on a close examination of figure \ref{fig:scan}, and on the
right-hand set of plots of figure \ref{fig:windowResurrections}.
Decreasing $H$ further, we also
observed interesting structures at the tip of the fractal tongue, near
to $H=-1$, with resonance windows merging to give rise to a pattern 
of half-rings on the phase diagram. These are shown in
the lower two plots in figure \ref{fig:saturn}. 

\begin{figure}
 \begin{center}
  \hspace*{-15pt}
\includegraphics[width=1.08\textwidth,angle=0]{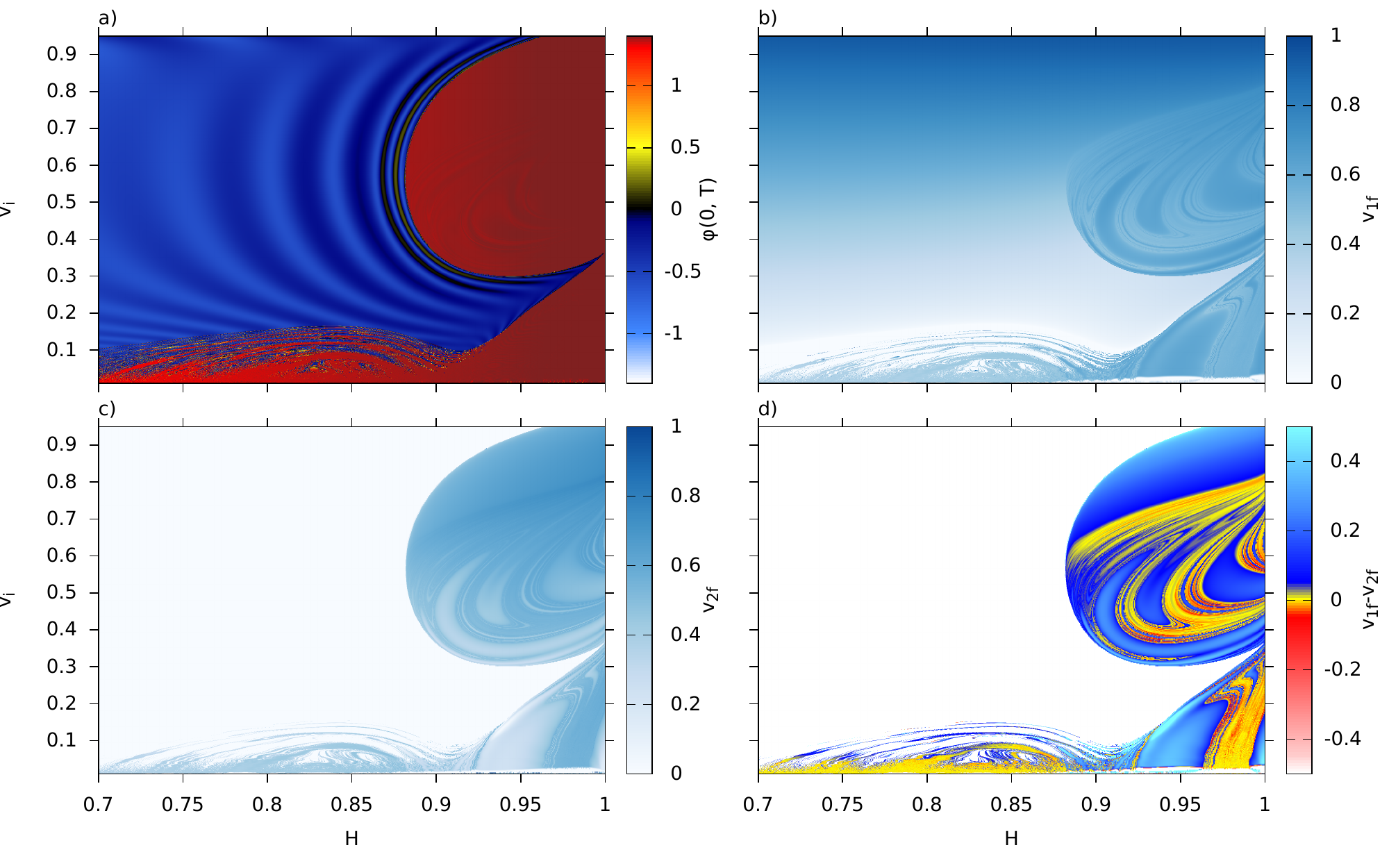}
 \vspace{-15pt}
 \caption{\label{fig:zoomscan}\small Antikink-boundary scattering at
large $H$.
(a): a zoomed-in view of figure \ref{fig:scan}
showing the value of the field at $x=0$ a time
$t_f=|x_0|/v_{i}+100$ after the start of the simulation; (b):
the measured final velocity of the reflected antikink; (c):
the measured final velocity of the emitted kink, if present, white
being plotted otherwise;
(d): the difference between these two
velocities.}
\end{center}
 \vspace{-7pt}
\end{figure}

For $H>0$, the
scattering can induce the metastable $\phi_1$ boundary to decay to
$\phi_3$, the true ground state,
with the creation of an extra kink.
This process, which has no analogue in the full-line theory, 
is visible in the extra red `kink' line in figure \ref{fig:kb}e.
The principal region of boundary decay occupies the solid red area on
the right edge of figure \ref{fig:scan}, and is examined in more
detail in figure \ref{fig:zoomscan}. 
If the boundary mode is sufficiently strongly excited by
the initial antikink impact, it behaves as an intermediate
state prior to the escape of a kink from
the boundary, analogous to the intermediate oscillon
state in the process of $K\bar K$ pair production on the full
line \cite{Manton:1996ex,Romanczukiewicz:2010eg,Romanczukiewicz:2005rm}.
Depending on their relative velocities, the reflected antikink and the
subsequently-emitted kink may appear separately
in the final state, or recombine to form a bulk oscillon.
Such collisions lead themselves to a fractal-like structure with
windows where the antikink and kink separate interspersed with regions of
oscillon production, just as in the
full-line theory (though with added complications due to 
interference with radiation from the boundary). Some of this structure
can be seen in figure \ref{fig:zoomscan}d, where the blue regions
inside the zone of boundary decay show windows of antikink and
kink separation, while the yellow regions correspond to the production
of a bulk oscillon, and also in the movies \texttt{M11} and \texttt{M12}.
Spacetime plots of some of the relevant processes, for $H=0.90$,
are shown in the right panels of figure \ref{fig:examples}:
scattering of the antikink with excitation of the boundary mode, 
but no kink production (d); production of
a separated $K\bar K$ pair, with the boundary decaying to the true
ground state (e); and recombination of the $K\bar K$ pair to form a bulk 
oscillon (f). 
A further intriguing feature of the region of boundary decay,
clearly visible in figures \ref{fig:scan} and \ref{fig:zoomscan}, is
the cusp-like nick, terminating at $(H,v_i) \approx
(1,0.365)$, which splits it into two disconnected parts. 
This appears to be associated with a
velocity-dependent vanishing of the effective
coupling between the incident antikink and the boundary mode. It
would be very interesting to find an analytical understanding of this
phenomenon, but we will leave this for future
work.

\section{Radiative decay of the boundary mode}
A significant feature of the  $\phi^4$ model is that its spectrum of
perturbative oscillations around the static kink or antikink solutions
contains an internal vibrational mode.
If the amplitude of the excitation is small enough and nonlinear
corrections can be neglected, this mode oscillates with almost-constant
amplitude $A$ and frequency $\omega_d = \sqrt{3}$.
For larger amplitudes nonlinearities start to play an important
role. It has been shown \cite{Manton:1996ex} that the first
anharmonic correction to the internal mode oscillation results in the
appearance of an outgoing wave with frequency $2\omega_d$, which is
above the mass threshold. 
The corresponding rate of
radiative energy loss is $dE/dt\sim A^4$, causing 
the mode to decay. The resulting
time dependence of the amplitude of the internal mode
follows the law $dA/dt\sim A^3$, where the explicit value of the
proportionality constant can be found using a Green's function
technique \cite{Manton:1996ex}.

For our boundary theory,
we have observed a similar pattern in the decay of small-amplitude
excitations of the boundary mode, but with a number of interesting new
features.
For small positive values of $H$, the frequency $\omega_B$ of the
linearised boundary mode, as predicted by
 (\ref{eq:kappa}), satisfies $2\omega_B>2$, and so the second harmonic of
this mode is able to propagate in the bulk\footnote{Recall that  
$m=2$ is the mass threshold for the bulk theory.}.
But as the boundary magnetic field $H$ increases, the frequency of the
boundary mode decreases, and when $H>H_2\approx 0.925$, 
$2\omega_B$ dips below $2$ and
the situation changes.
The second harmonic can no longer propagate into the bulk,
and this channel of radiative energy
loss from the boundary is terminated.  Only the next harmonic, which
appears in the third order of the perturbation series, can be seen in the
power spectrum.  The radiation loss rate becomes $dE/dt\sim A^6$
and the decay rate is reduced to $dA/dt\sim A^5$.

The situation changes again as $H$ increases beyond
$H=H_3\approx 0.982$, when $3\omega_B$ falls below $2$
and the third harmonic joins the second, trapped below
the mass threshold.
Theoretically, as $H \to 1$ and $\omega_B \to 0$ this pattern will repeat
an infinite number of times, so that whenever $\frac{2}{n+1}<\omega_B
< \frac{2}{n}$, 
the amplitude of the decaying mode should satisfy, to leading order,
the equation $dA/dt\sim A^{2n-1}$.

\begin{figure}
 \begin{center}
   \includegraphics[width=0.95\textwidth,angle=0]{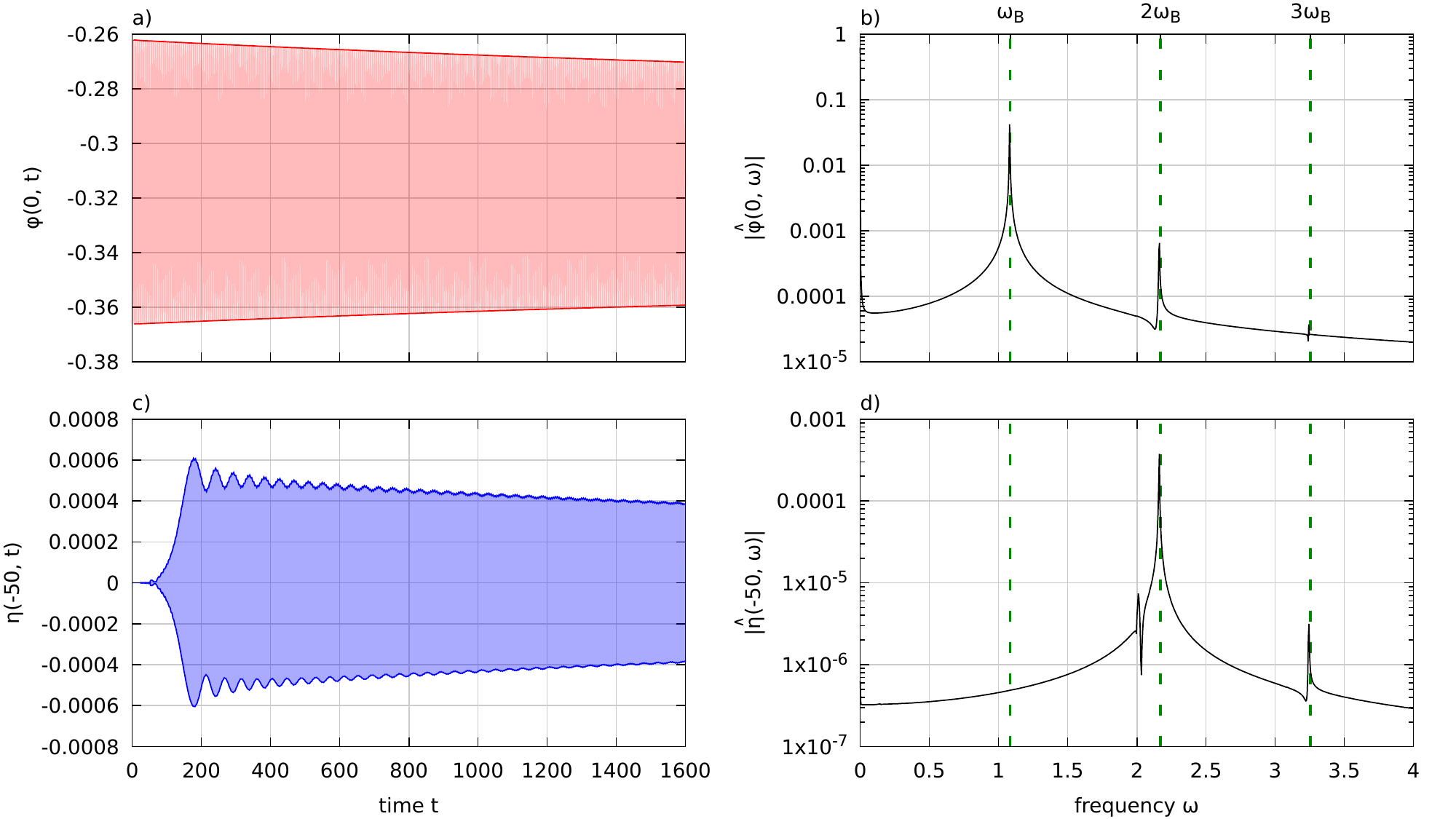}
 \vspace{-7pt}
 \caption{\label{fig:belowH2}\small Evolution of the boundary mode for
$H=0.90<H_2$, with
initial conditions $\phi(x,0)=\phi_1(x)+0.05\,\eta_B(x)$,
$\phi_t(x,0)=0$\,:
a) the amplitude of the boundary
mode as a function of time; b) its power spectrum, found 
by taking the Fourier transform of $\phi(0,t)$ for $0<t<1600$; c)
the values of $\eta(x,t):=\phi(x,t)-\phi_1(x)$ at $x=-50$; 
d) its power spectrum, taken from $\eta(-50,t)$ 
for $200<t<1600$.
In plots
 a) and c), solid non-transparent lines join local extrema of the
measured field at the given position.}
\end{center}
 \vspace{-7pt}
\end{figure}
\begin{figure}[h]
 \begin{center}
   \includegraphics[width=0.95\textwidth,angle=0]{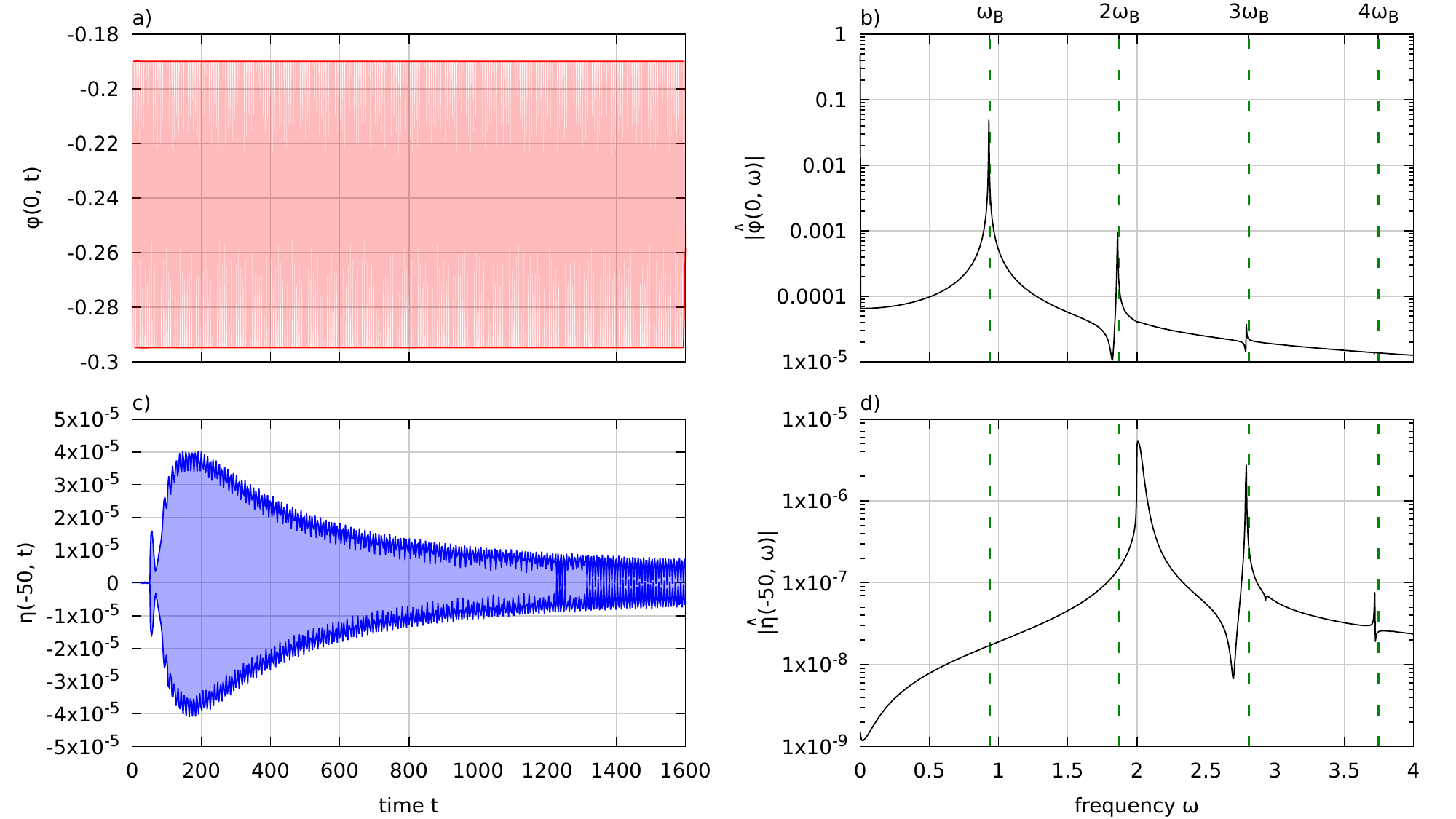}
 \vspace{-7pt}
 \caption{\label{fig:aboveH2}\small 
The same sequence of plots as in figure \ref{fig:belowH2}, but now for
$H=0.94>H_2$\,. }
\end{center}
\end{figure}

Figure \ref{fig:belowH2} shows the behaviour of the field on the boundary and 
at $x=-50$, in the far field zone,
with initial conditions 
$\phi(x,0)=\phi_1(x)+0.05\,\eta_B(x)$, $\phi_t(x,0)=0$ and $H=0.90<H_2$.
The power spectrum of the field on the boundary is dominated by
boundary mode oscillating with the theoretically
predicted frequency $\omega_B=1.08509$.
There are also two peaks at $2\omega_B$ and $3\omega_B$.  Since
$\omega_B<m$, this lowest mode cannot propagate and indeed, there is no
trace of it in the far field zone. The mode with the frequency
$2\omega_B$ is already in the scattering
spectrum, so this mode does propagate, causing the energy loss from
the boundary mode, as seen in figure \ref{fig:belowH2}a.

The picture is different when $H=0.94>H_2$ (see
figure~\ref{fig:aboveH2}). The mode with frequency
$\omega_B=0.93643$ still dominates the power spectrum of the boundary
excitations, but its decay is much slower, reflecting the fact that the 
mode $2\omega_B$ is now below the mass
threshold and cannot propagate into the bulk.
As can be seen from  the power spectrum in the far field zone plotted
in figure~\ref{fig:aboveH2}d, the radiation is much less than in the
previous case.
There are two dominant frequencies,
$\omega=2$ and $\omega=3\omega_B$. The presence of the peak at
$3\omega_B$ is natural, since this is the
first harmonic above the mass threshold. The peak at $\omega=2$
originates from near-threshold bulk modes, excited by the initial
conditions via the nonlinearities, which disperse only 
slowly away from the boundary \cite{Bizon:2011pb}.

\begin{figure}
 \begin{center}
   \includegraphics[width=0.8\textwidth,angle=0]{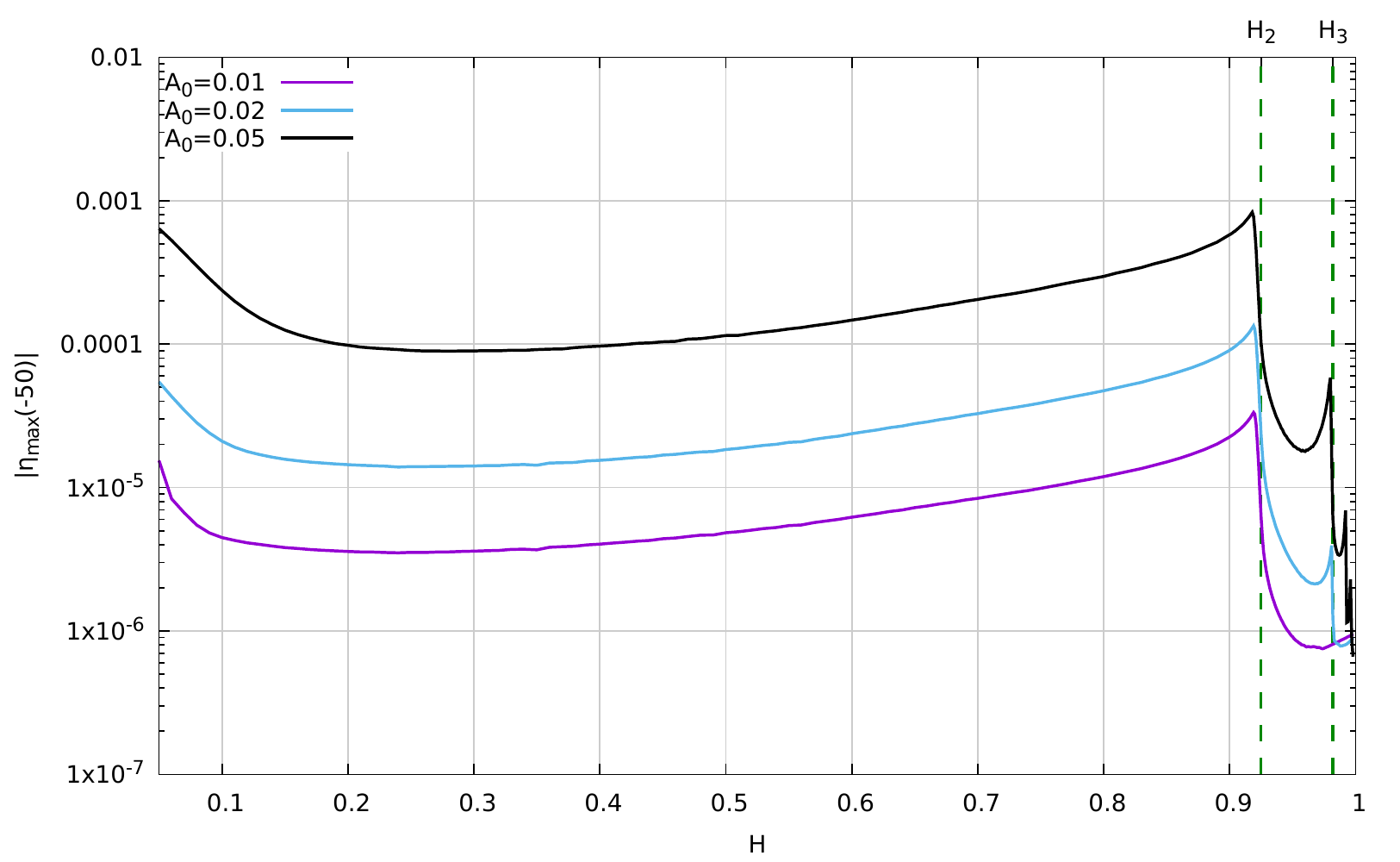}
 \vspace{-10pt}
 \caption{\label{fig:radiation}\small Maximal radiation amplitude 
$|\eta_{\rm max}(x)|$, where $\eta(x,t)=\phi(x,t)-\phi_1(x)$ is 
the deviation of the field
from its static value, at  $x=-50$ for the kicked initial conditions
 $\phi(x,0)=\phi_1(x)$, $\phi_t(x,0)=A_0\omega_B\eta_B(x)$ as a
function of $H$, for three different values of 
$A_0$.}
\end{center}
\vspace{-3pt}
\end{figure}
\begin{figure}[h]
 \begin{center}
   \includegraphics[width=0.8\textwidth,angle=0]{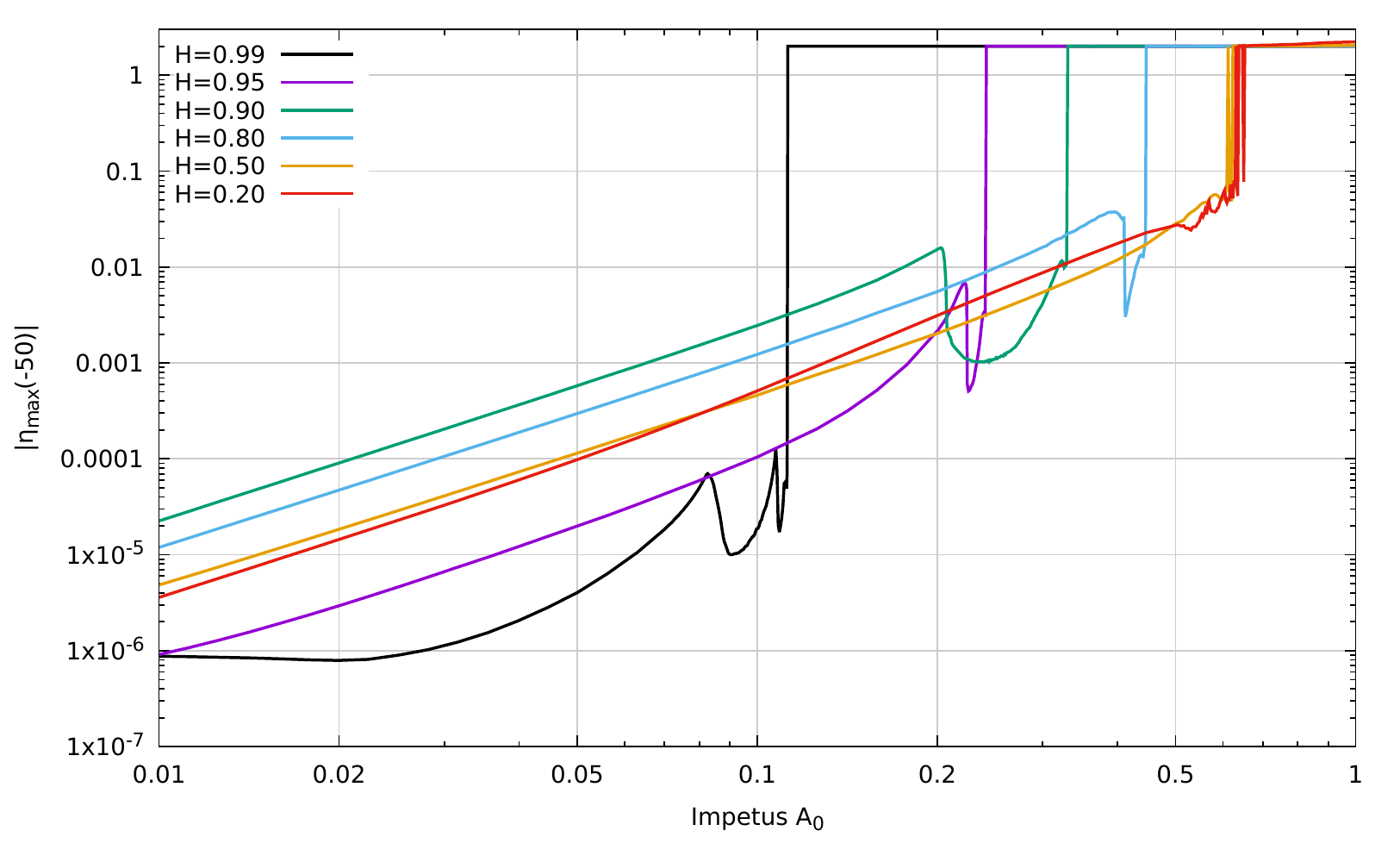}
 \vspace{-10pt}
 \caption{\label{fig:radAmp}\small A log-log plot of the maximal 
radiation amplitude at $x=-50$
for the kicked initial conditions $\phi(x,0)=\phi_1(x)$,
$\phi_t(x,0)=A_0\omega_B\eta_B(x)$
as a function of $A_0$.
Note that for small $A_0$ and $H<H_2$ the power law decay rate is universal. }
\end{center}
 \vspace{-10pt}
\end{figure}

Another test of the scenario is to consider the radiation from a
``kicked'' boundary initial condition $\phi(x,0)=\phi_1(x)$,
$\phi_t(x,0)=A_0\omega_B\eta_B(x)$ in the far field zone.
Our numerical results for this case
are presented in figures~\ref{fig:radiation} and \ref{fig:radAmp}.

Figure~\ref{fig:radiation} shows the $H$-dependence of maximal amplitude
of the field measured at $x=-50$, far away from the boundary, for
three small values of the initial impetus $A_0$ given to 
the boundary mode.
Note that the radiation amplitude drops sharply when $H$ crosses
$H_2$ and $H_3$, as predicted by our general considerations.

Figure \ref{fig:radAmp} shows a log-log plot of
the dependence of the maximal amplitude at $x=-50$ on $A_0$.
For small values of $A_0$ and $H<H_2$, all curves have
the same slope, fitting the expected
$\sim A_0^2$ dependence.
The curve for $H=0.95$ shows a significant reduction in the radiation
amplitude, reflecting the loss
of a decay channel as $H$ passes $H_2$. However its slope for small
values of $A_0$ appears to be relatively unchanged from that of the
previous curves, even though our
previous considerations based on the propagation of the third harmonic
would suggest an $\sim A_0^3$ dependence.
It may be that slow (near-threshold) bulk modes, visible in
figure~\ref{fig:aboveH2}d in the peak at $\omega=2$,
are obscuring the effect we are looking for. It is possible
that this could be tested by waiting significantly longer before measuring 
the radiation, to allow the slow modes to die away, but a
more-detailed study would be needed to draw a clear conclusion.

Another interesting feature visible on each curve is that as 
$A_0$ reaches some (curve-dependent)
critical value, the
radiation flux suddenly dips.  As will be discussed in the next
section, this effect is associated with the nonlinear effect of the
reduction in the frequency
of the boundary mode with increasing amplitude.

Finally, for even larger values of the intial impetus
we can see a large increase of the amplitude of the
field in the far zone.
This is a signature of a non-perturbative effect, the excitation
at the boundary
becoming strong enough to destabilise it completely, with the emission
of a kink into the bulk flipping the field there into the other vacuum. 
Some further observations concerning this phenomenon
are reported in section~\ref{creation} below.

\section{Higher-order nonlinear effects and amplitude-dependent decay rates}
\label{sec:higherorder}
In the last section we principally considered boundary mode decay in the
small-amplitude regime where the boundary mode itself could be treated
linearly.
For larger amplitudes the
frequency of the mode's oscillation is 
lowered, just  as in the case of an
anharmonic oscillator or the simple pendulum. Numerical simulations 
of the oscillations of a full-line kink \cite{Manton:1996ex}
also exhibit this behaviour,
which is typical for many nonlinear systems.

In the evaluation of the critical values $H_n$  above, we implicitly
assumed that the amplitude of the excitation was small, so that
its frequency was that predicted by the linearised equations.
However for larger amplitudes, given the amplitude-dependent frequency
reduction just described, it is possible that even for
$H<H_2$, the actual frequency of the boundary
mode, $\tilde{\omega}_B$, will
be lower than $m/2$. Then
the decay rate will be slower than that observed
for smaller amplitudes, since the
second harmonic will not couple directly with any propagating bulk modes.
However, the amplitude of the boundary mode will decrease with time
due to the outgoing radiation, causing its frequency to grow.
Provided $H<H_2$, once the amplitude has decreased far enough,
the second harmonic will enter the scattering spectrum. 
In such a case we can expect to
observe an intriguing phenomenon: while initially the radiation flux from
the boundary is relatively small and the decay rate rather slow,
after some time there will be a sudden increase
of the radiation flux and a switch to a much faster decay rate.

\begin{figure}
 \begin{center}
   \includegraphics[width=1\textwidth,angle=0]{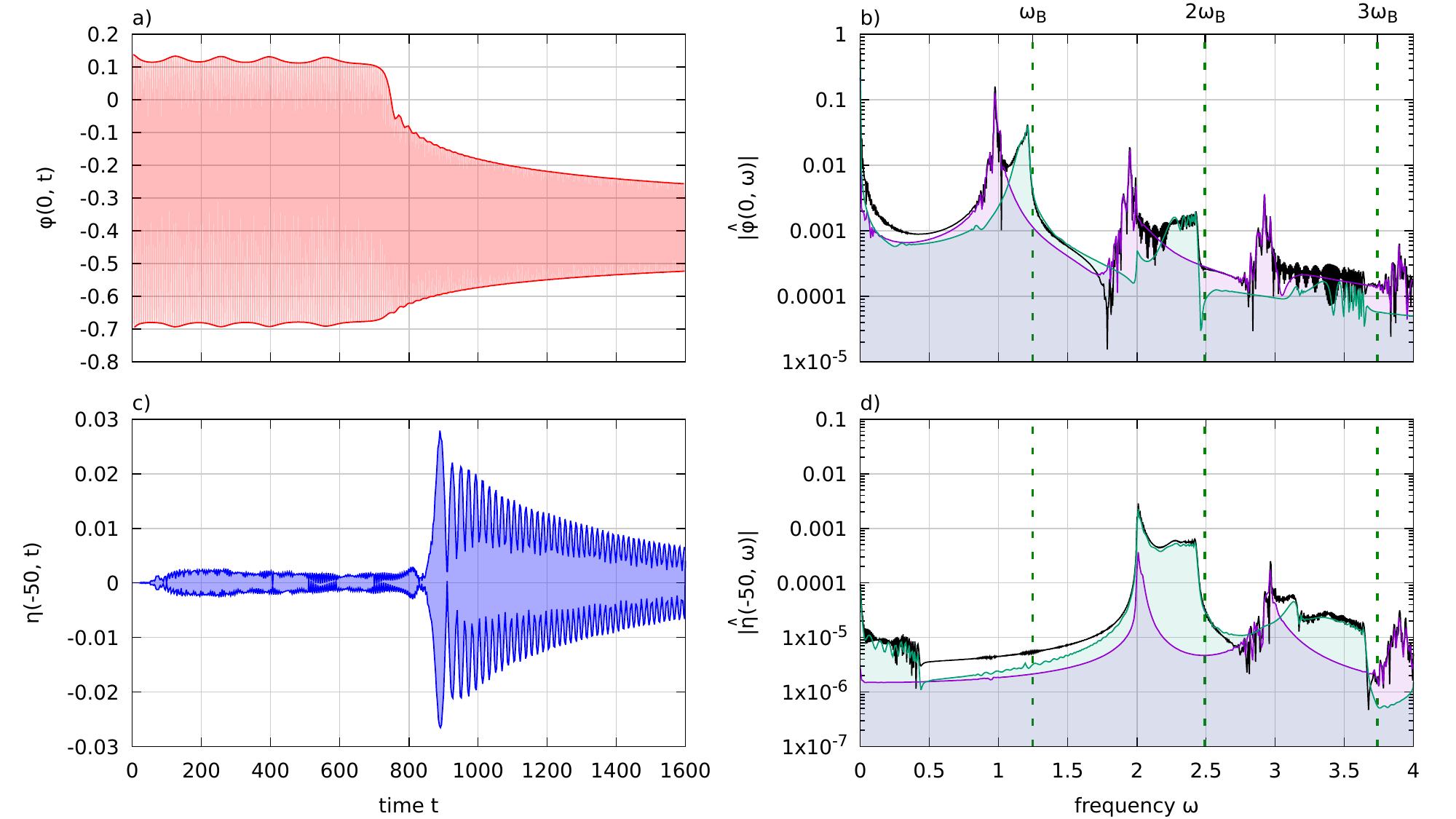}
 \caption{\label{fig:intrigue}\small Evolution of the boundary mode
for $H=0.8393<H_2$ and large initial amplitude
$A_0=0.3$. Plots a) and c) show the
 time evolution of the amplitudes of the boundary mode and radiation
field respectively, 
from the values of the field at $x=0$ and $x=-50$.
Plots b) and d) show power spectra at the positions $x=0$ and $x=-50$.
The black lines show the power spectra from $\phi(0,t)$ for times
$0<t<1600$ (plot b)) and $\phi(-50,t)$ for times $200<t<1600$ (plot d)). 
The purple and green lines and filled areas
show the power spectra for times before and after the transition 
($0<t<750$ and $750<t<1600$ for $x=0$; $200<t<800$ and $800<t<1600$ for 
$x=-50$).  }
\end{center}
\end{figure}

Numerical work confirms that this effect really exists,
as can be seen in figure \ref{fig:intrigue} and movie \texttt{M13}, 
which show the decay of the amplitude of the boundary
mode.  For about the first 750 units of time the amplitude
changes very slowly, albeit with a small modulation, after 
which there is a sudden transition to a much more
rapid decay.
In the far field zone this effect can be observed as a
sudden jump of the radiation flux, by about one order of magnitude.

In the power spectrum plotted in figure \ref{fig:intrigue}d one can
clearly see  a large peak just below $\omega= 1$, which is the initial
frequency of the mode. While the amplitude slowly decreases the
frequency grows until it crosses 1, after which point the decay runs much
faster. We can also see a drift of the frequency up to
$\omega_B=1.24666$.

This slow-then-fast behaviour is reminiscent of higher-dimensional
oscillon decay.
In \cite{Copeland:1995rm,Honda:2002ar,Fodor:2006hg}
it was observed that oscillons 
in two and three spatial dimensions lose their
energy very slowly for tens of thousands of
oscillations until they reach some critical frequency, above
which they quickly decay to the vacuum.

\section{Creation of kinks from an excited boundary}
\label{creation}
The final phenomenon we investigated was the creation of kinks from the
metastable boundary. We previously observed that this could be induced
in certain scattering processes at large $H$. To view it in
isolation, we instead excited the boundary mode directly, taking
initial conditions of the two types (``stretched'' and ``kicked") used
earlier.
First, we used initial condition
\begin{equation}
 \phi(x,0)=\phi_1(x)+A_0\eta_B(x),\;\;\;\;\phi_t(x,0)=0
\end{equation}
with $\eta_B(x)$ the boundary profile for the linearised prolem, as
an approximation to the boundary mode 
at its largest deviation from equilibrium; and second, we took
\begin{equation}
 \phi(x,0)=\phi_1(x),\;\;\;\;\phi_t(x,0)=A_0\omega_B\eta_B(x)
\end{equation}
representing the ``kicked'' boundary.
As before, we normalized the profile of the boundary mode in such a way
that $\eta_B(0)=1$.

\begin{figure}
 \begin{center}
   \includegraphics[width=0.8\textwidth,angle=0]{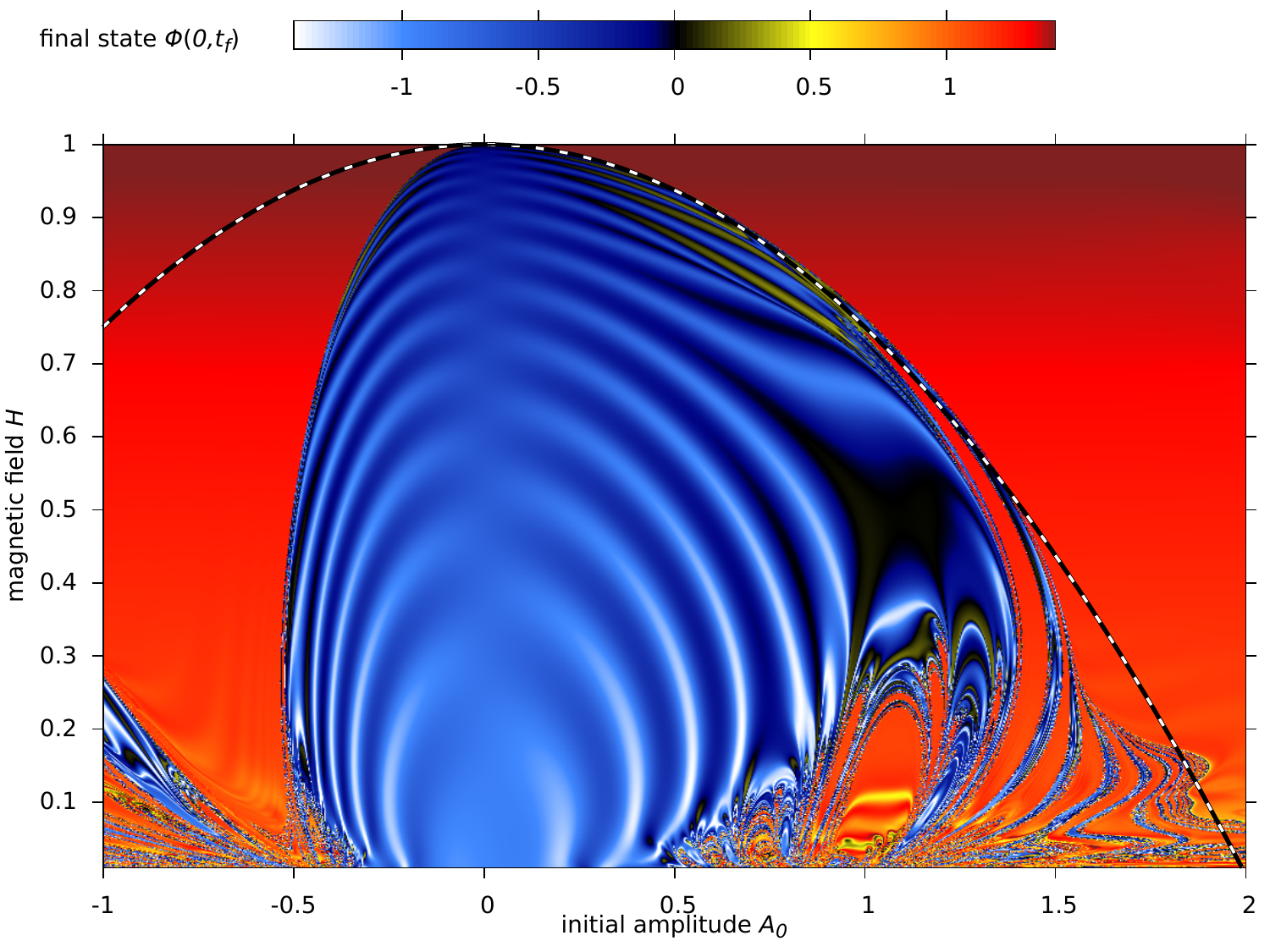}
 \caption{\label{fig:creationScan}\small The field value on the 
boundary at time $t_f=50$ for initial conditions
 $\phi(x,0)=\phi_B(x)+A_0\eta_B(x)$, $\phi_t(x,0)=0$. The blue colour
represents the region without kink creation. The dashed line 
corresponds to the function $H=1-A^2/4$.}
\end{center}
 \vspace{-20pt}
\end{figure}

\begin{figure}
 \begin{center}
   \includegraphics[width=0.8\textwidth,angle=0]{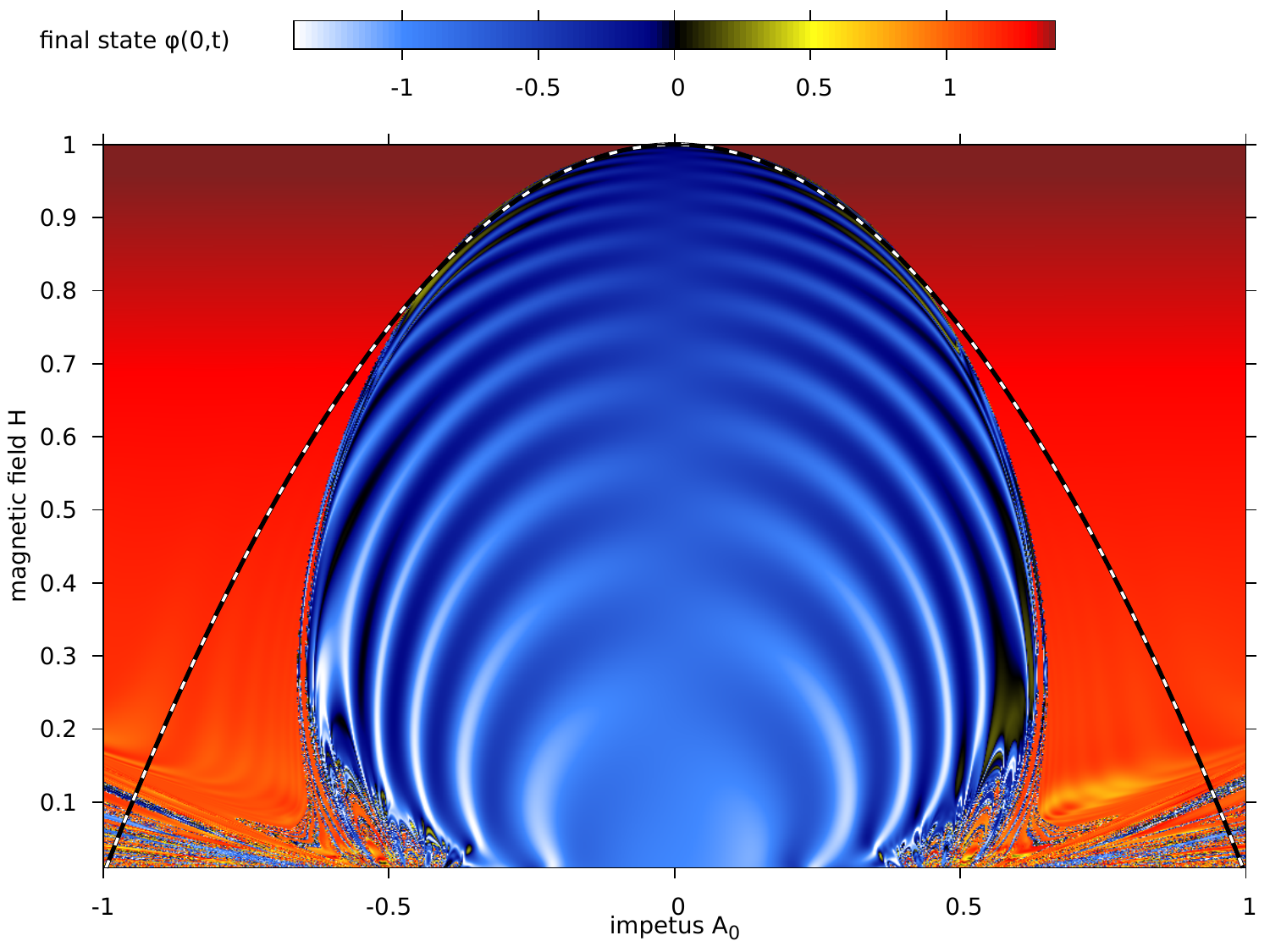}
 \caption{\label{fig:creationKick}\small The field value on the
boundary at time $t_f=50$ for initial conditions $\phi(x,0)=\phi_B(x)$,
 $\phi_t(x,0)=A_0\omega_B\eta_B(x)$. The blue colour represents the region 
without kink creation. The dashed line corresponds to the function
$H=1-A^2$.}
\end{center}
 \vspace{-20pt}
\end{figure}

In both cases, if $A_0$ is taken to be sufficiently small,
the boundary oscillates with frequency $\omega_B$ and the
amplitude $A_0$.
However, as $A_0$ becomes larger, the nonlinear processes discussed
above start to play a significant role and further, as
the initial energy of the excited mode becomes sufficient, 
outgoing kinks can be observed
in the far zone, as seen in 
figures~\ref{fig:creationScan} and \ref{fig:creationKick}.

Note that for large $H\to 1$ the boundary mode profile resembles the difference between the unstable boundary solution
(a saddle point of the energy) and the stable boundary solution:
\begin{equation}
 \eta_B(x)\approx\frac{\tanh(x+X_0)-\tanh(x-X_0)}{2\tanh(X_0)}
\end{equation}

Therefore the boundary mode, with appropriate amplitude, being added to the static boundary solution
yields the unstable boundary. When the solution crosses the saddle point of energy  it decays into another static solution with
 an additional kink is emitted from the boundary.

Therefore 
the critical value of the amplitude of the boundary mode  
for the production of the kinks is:
\begin{equation}
 A_{crit} = \phi_2-\phi_1=2\sqrt{1-H}.
\end{equation}
This  critical amplitude is in very good
 agreement with the first type of the initial conditions for positive values of $A_0$.
Only for very small values of $A_0$ is there a symmetry $A_0\to-A_0$.
For larger $A_0>0$ the excitation have less energy than the
excitation for $-A_0$. Therefore the critical line for kink creation, from
the left side of the plot, is much closer to the centre ($A_0=0$).

For initial conditions of the second type, the energy for 
$A_0$ and $-A_0$ is exactly the same and therefore the plots
look much more symmetric.  For $H\to1$ the critical amplitude is almost exactly
$A_{crit}=\sqrt{1-H}$, half as big as in the first case.

\section{Conclusions}
Our investigations of the boundary $\phi^4$ theory
have shown that it offers a considerably richer variety of
resonance phenomena than the bulk theory, within a setting
where analytical progress can be made. Key features include the
modification of the force leading to the sharpening of window
boundaries and the new critical velocity $v_{cr}$, the resurrection
of the first `missing' scattering window, the observation of the boundary
oscillon, and the collision-induced decay of the metastable boundary
vacuum for $H$ near to $1$. Much of our work has been numerical and 
many issues remain for further study, the
most pressing being the development of a reliable
moduli space approximation incorporating the boundary degrees of
freedom (see \cite{Antunes:2003kh} for some earlier work on this
issue). This model is sufficiently
simple that it should offer the ideal playground for the development of
better analytical techniques for the
understanding of more general nonintegrable field
theories.

\smallskip

\section*{Acknowledgements}
We would like to thank  Piotr Bizo\'n,  Robert Parini and Wojtek
Zakrzewski for
discussions. The work of PED was supported by an STFC Consolidated Grant,
ST/L000407/1, and by the GATIS Marie Curie FP7 network (gatis.desy.eu)
under REA Grant Agreement No 317089.
AH thanks the BSU Student Grant Program for support, and
YS thanks the Russian Foundation for Basic Research (Grant No. 16-52
-12012), DFG (Grant LE 838/12-2) and JINR Bogoljubov-Infeld Programme.


\appendix

\section{Numerical methods}
\label{app:methods}
In this Appendix we describe some details of the numerical methods
used in our simulation.
In our numerical code we used the following discretization:
\begin{equation}
 u_n = \phi(-nh),\;\;n=0\ldots N\,.
\end{equation}
To calculate spatial derivatives we used a
fourth-order central 
difference scheme 
\begin{equation}
  D^2u_n = \frac{1}{12h^2} \left(-u_{n-2}+16u_{n-1}-30u_n+16u_{n+1}-u_{n+2}\right)
\end{equation}
for all points far enough from the boundary, $n\leq 2$.
This scheme can be derived using Lagrange polynomial approximation:
\begin{equation}
 u(x) = \sum_{i=n-m}^{n+m}u_i\ell_i(x),\;\;\ell_i(x) = \prod_{\begin{smallmatrix}j=n-m\\ j\neq i\end{smallmatrix}}^{n+m}\frac{x-jh}{ih-jh}.
\end{equation}
However for the two points closest to the boundary we have to use a
different basis
\begin{equation}\label{eq:approx}
 u(x) = H\bar\ell_0(x)+\sum_{i=0}^3u_i\tilde\ell_i(x),
\end{equation}
where
\begin{align}
 \tilde \ell_{i>0}(x) &= \frac{x^2}{(ih)^2}\prod_{\begin{smallmatrix}j=1\\ j\neq i\end{smallmatrix}}^3\frac{x-jh}{ih-jh},\\
 \tilde \ell_{0}(x) &= \left(1+\frac{11}{6}x\right)\prod_{j=1}^3\frac{x-jh}{-jh},\\
 \bar \ell_{0}(x) &= x\prod_{j=1}^3\frac{x-jh}{-jh}.
\end{align}
Note that for points within the interpolation intervals
\begin{equation}
 \ell_i(x_j)=\tilde\ell_i(x_j)=\delta_{ij},\;\;\bar\ell(x_i)=0,\;\;\tilde\ell_i'(0)=0,\;\;\bar\ell'(0) = 1.
\end{equation}
The above relations prove that formula (\ref{eq:approx}) really interpolates the function with appropriete boundary condition
\begin{equation}
 u(nh) =u_n,\;\;n\leq 3\;\;\text{and}\;\;u'(0) = H.
\end{equation}
From this approximation it is straightforward to calculate the second derivative for the first two points:


\begin{align}
 D^2u_0 &= -\frac{1}{18h^2}\left(85u_0+66hH-108u_1+27u_2-4u_3\right),\\
 D^2u_1 &= \frac{1}{18h^2}\left(29u_0+6hH-54u_1-2u_3+27u_2\right).
\end{align}

\section{Supplementary material}
\label{app:supplementary}
We have prepared a number of short movies, labelled \texttt{M01} \dots
\texttt{M13}, to illustrate aspects of
our findings, which are listed in this appendix. The movies themselves
can be found at~\cite{animations}.
\newcommand\movitem[1]{\item\hspace*{-20pt}\texttt{\href{https://arxiv.org/src/1508.02329/v2/anc/#1}{#1}}\,:\,}
\medskip

\noindent
The first six movies show the processes depicted in figure
\ref{fig:examples} a -- f\,:
\begin{enumerate}[label=]
\movitem{M01\_BndryScattering\_Hminus040\_v020.mov} 
$H=-0.4$, $v=0.20<v_{cr}(H)$\\
Almost-perfect reflection of the incident antikink, which has insufficient
energy to get over the saddle-point potential barrier.
\movitem{M02\_BndryScattering\_Hminus040\_v0333.mov} 
$H=-0.4$, $v=0.333=v_{cr}(H)$\\
Antikink incident
at the critical velocity, leading to the creation of the saddle-point
configuration. Note, this movie (and the associated figure
\ref{fig:examples}b) is somewhat idealised, as in practice it is
impossible to tune the initial velocity finely enough to hit
the true critical velocity precisely. Instead, we patched together 
an animation up to
$t=40$ with the static solution thereafter.
\movitem{M03\_BndryScattering\_Hminus040\_v040.mov}
$H=-0.4$, $v=0.40>v_{cr}(H)$\\
Single bounce, with subsequent escape of the antikink.
Note that the acceleration of the antikink after it surpasses the
potential barrier is clearly visible.
\movitem{M04\_BndryScattering\_Hplus090\_v035.mov}
$H=0.9$, $v=0.35$\\
Single bounce, with excitation of both the $H>0$ boundary mode and the
internal mode of the antikink.
\movitem{M05\_BndryScattering\_Hplus090\_v037.mov}
$H=0.9$, $v=0.37$\\
Single bounce exciting the boundary mode strongly enough to induce decay 
of the metastable boundary state, creating an additional kink in the
bulk.
\movitem{M06\_BndryScattering\_Hplus090\_v039.mov}
$H=0.9$, $v=0.39$\\
A similar process to \texttt{M5}, but here
the relative velocities of the emitted
kink and antikink are such that a bulk oscillon is formed instead
of a separated kink-antikink pair.
\end{enumerate}
\medskip

\noindent
The next six movies scan through a range of velocities at
constant $H$. Movies \texttt{M07} -- \texttt{M10} show 
`traditional' window formation as in the full-line case (equivalent to
\texttt{M09}).  The sharpened edges of the windows
for $H<0$, caused by the presence there of a potential barrier, are
visible on careful comparision of \texttt{M7} and \texttt{M8} (for
$H<0$) with \texttt{M9} and \texttt{M10} (for $H\ge 0$). For movies
\texttt{M11} and  \texttt{M12}, $H$ is in the region where
collision-induced boundary decay is possible.
\begin{enumerate}[label=]
\setcounter{enumi}{6}
\movitem{M07\_VelocityScan\_Hminus040.mov} $H=-0.4$
\movitem{M08\_VelocityScan\_Hminus020.mov} $H=-0.2$
\movitem{M09\_VelocityScan\_H000.mov} $H=0$
\movitem{M10\_VelocityScan\_Hplus020.mov} $H=0.2$
\movitem{M11\_VelocityScan\_Hplus090.mov} $H=0.9$
\movitem{M12\_VelocityScan\_Hplus095.mov} $H=0.95$
\end{enumerate}
\medskip

\noindent
Finally, movie \texttt{M13} shows the slow-then-fast relaxation of the
boundary mode. Four plots are shown: on the left, the field values
next to, and slightly further from, the boundary; on the right, the
amplitude of the field at the boundary, and an estimate $\omega(0,t)
:=2\pi/T$ of its instantaneous frequency, where $T$ is the time
between successive minima of $\phi(0,t)$.
\begin{enumerate}[label=]
\setcounter{enumi}{12}
\movitem{M13\_Relaxation\_Hplus08393\_A030.mov} $H=0.8393$, $A_0=0.3$
\end{enumerate}

\begin{small}

\end{small}


\begin{thebibliography}{99}
\bibitem{Kondo}
 P.~Fendley,
  ``Kinks in the Kondo problem'',
  Phys.\ Rev.\ Lett.\ {\bf 71} (1993) 2485\\
 doi:10.1103/PhysRevLett.71.2485
  [cond-mat/9304031].
\bibitem{Olsen:1981a}
 O.H.~Olsen and M.R.~Samuelsen,
  ``Fluxon propagation in long Josephson junctions with external
  magnetic field'',
  J.\ Appl.\ Phys.\ {\bf 52} (1981) 6247\\
doi:http://dx.doi.org/10.1063/1.328567.
\bibitem{Alcaraz:1987}
 F.~Alcaraz, M.~Barber, M.~Batchelor, R.~Baxter and G.~Quispel,
  ``Surface exponents of the quantum XXZ, Ashkin-Teller and Potts
  models'',
  J.\ Phys.\ A {\bf 20} (1987) 6397\\
doi:10.1088/0305-4470/20/18/038.
\bibitem{Kane:1992zza}
  C.L.~Kane and M.P.A.~Fisher,
  ``Transmission through barriers and resonant tunneling in an
  interacting one-dimensional electron gas'',
  Phys.\ Rev.\ B {\bf 46} (1992) 15233\\
doi:10.1103/PhysRevB.46.15233.
\bibitem{Ghoshal:1993tm}
  S.~Ghoshal and A.B.~Zamolodchikov,
  ``Boundary S matrix and boundary state in two-dimensional integrable
  quantum field theory'',
  Int.\ J.\ Mod.\ Phys.\ A {\bf 9} (1994) 3841
   [Erratum-ibid.\ A {\bf 9} (1994) 4353]\\
doi:10.1142/S0217751X94001552
  [hep-th/9306002].
\bibitem{Bowcock:1995vp}
  P.~Bowcock, E.~Corrigan, P.E.~Dorey and R.H.~Rietdijk,
  ``Classically integrable boundary conditions for affine
  Toda field theories'',
  Nucl.\ Phys.\ B {\bf 445} (1995) 469\\
doi:10.1016/0550-3213(95)00153-J
  [hep-th/9501098].
\bibitem{Rubakov:1981rg}
  V.A.~Rubakov,
  ``Superheavy magnetic monopoles and proton decay'',
  JETP Lett.\  {\bf 33} (1981) 644
  [Pisma Zh.\ Eksp.\ Teor.\ Fiz.\  {\bf 33} (1981) 658].
\bibitem{Wen:1990se}
  X.G.~Wen,
  ``Chiral Luttinger liquid and the edge excitations in the fractional
  quantum Hall states'',
  Phys.\ Rev.\ B {\bf 41} (1990) 12838\\
doi:10.1103/PhysRevB.41.12838.
\bibitem{Antunes:2003kh}
  N.D.~Antunes, E.J.~Copeland, M.~Hindmarsh and A.~Lukas,
  ``Kink boundary collisions in a two-dimensional scalar field
  theory'',
  Phys.\ Rev.\ D {\bf 69} (2004) 065016\\
doi:10.1103/PhysRevD.69.065016
  [hep-th/0310103].
\bibitem{Campbell:1983xu}
  D.K.~Campbell, J.F.~Schonfeld and C.A.~Wingate,
  ``Resonance structure in kink - antikink interactions in $\phi^{4}$
  theory'',
  Physica {\bf 9D} (1983) 1\\
doi:10.1016/0167-2789(83)90289-0.
\bibitem{Peyrard:1984qn}
  M.~Peyrard and D.K.~Campbell,
  ``Kink antikink interactions in a modified sine-Gordon model'',
  Physica {\bf 9D} (1983) 33\\
doi:10.1016/0167-2789(83)90290-7.
\bibitem{Anninos:1991un}
  P.~Anninos, S.~Oliveira and R.A.~Matzner,
  ``Fractal structure in the scalar $\lambda(\phi^2-1)^2$ theory'',
  Phys.\ Rev.\  D {\bf 44} (1991) 1147\\
doi:10.1103/PhysRevD.44.1147.
\bibitem{Goodman:2007}
  R.~Goodman and R.~Haberman,
  ``Chaotic scattering and the n-bounce resonance in solitary-wave
  interactions'',
  Phys.\ Rev.\ Lett.\ {\bf 98} (2007) 104103\\
doi:10.1103/PhysRevLett.98.104103 [arXiv:nlin/0702048].
\bibitem{Dorey:2011yw}
  P.~Dorey, K.~Mersh, T.~Romanczukiewicz and Y.~Shnir,
  ``Kink-antikink collisions in the $\phi^6$ model'',
  Phys.\ Rev.\ Lett.\  {\bf 107} (2011) 091602\\
doi:10.1103/PhysRevLett.107.091602
  [arXiv:1101.5951 [hep-th]].
\bibitem{Arthur:2015mva}
  R.~Arthur, P.~Dorey and R.~Parini,
  ``Breaking integrability at the boundary: the sine-Gordon model
  with Robin boundary conditions'',
  J.\ Phys.\ A {\bf 49} (2016) no.16,  165205\\
  doi:10.1088/1751-8113/49/16/165205
  [arXiv:1509.08448 [hep-th]].
\bibitem{Bogolyubsky:1976nx}
  I.L.~Bogolyubsky and V.G.~Makhankov,
  ``On the pulsed soliton lifetime in two classical relativistic
  theory models'',
  JETP Lett.\  {\bf 24} (1976) 12.
\bibitem{animations}
 Animations of various aspects of the model can be found at
\href{https://arxiv.org/src/1508.02329/v2/anc/}{\texttt{https://arxiv.org/src/1508.02329/v2/anc/}.}
\bibitem{Romanczukiewicz:2010eg}
  T.~Romanczukiewicz and Y.~Shnir,
  ``Oscillon resonances and creation of kinks in particle
  collisions'',
  Phys.\ Rev.\ Lett.\  {\bf 105} (2010) 081601\\
doi:10.1103/PhysRevLett.105.081601
  [arXiv:1002.4484 [hep-th]].
\bibitem{Sugiyama:1979mi}
  T.~Sugiyama,
  ``Kink - antikink collisions in the two-dimensional $\phi^4$ model'',
  Prog.\ Theor.\ Phys.\  {\bf 61} (1979) 1550\\
  doi:10.1143/PTP.61.1550.
\bibitem{Manton:1996ex}
  N.S.~Manton and H.~Merabet,
  ``$\phi^4$ kinks - gradient flow and dynamics'',
  Nonlinearity {\bf 10} (1997) 3\\
doi:10.1088/0951-7715/10/1/002
  [hep-th/9605038].
\bibitem{Romanczukiewicz:2005rm}
  T.~Romanczukiewicz,
  ``Creation of kink and antikink pairs forced by radiation'',
  J.\ Phys.\ A  {\bf 39} (2006) 3479\\
doi:10.1088/0305-4470/39/13/022
  [hep-th/0501066].
\bibitem{Bizon:2011pb}
P.~Bizo\'n, T.\ Chmaj and N.\ Szpak,
``Dynamics near the threshold for blowup in the one-dimensional
focusing nonlinear Klein-Gordon equation'',
J.\ Math.\ Phys.\ \textbf{52}, (2011) 103703\\
doi:10.1063/1.3645363.
\bibitem{Copeland:1995rm}
E.J.~Copeland, M.~Gleiser and H.R.~Mueller,
``Oscillons: resonant configurations during bubble collapse'',
Phys.\ Rev.\ D {\bf 52} (1995) 1920\\
doi:10.1103/PhysRevD.52.1920
  [hep-ph/9503217].
\bibitem{Honda:2002ar}
E.P.~Honda and M.W.~Choptuik,
``Fine structure of oscillons in the spherically symmetric $\phi^4$
Klein-Gordon model''
Phys.\ Rev.\ D \textbf{65} (2002) 084037\\
doi:10.1103/PhysRevD.65.084037
  [hep-ph/0110065].
\bibitem{Fodor:2006hg}
G.\ Fodor, P.\ Forg\'acs, P.\ Grandcl\'ement and I.\ R\'acz,
``Oscillons and quasi-breathers in the $\phi^4$ Klein-Gordon model''
Phys.\ Rev.\ D \textbf{74} (2006) 124003\\
doi:10.1103/PhysRevD.74.124003
  [hep-th/0609023].


\end{thebibliography}
\end{document}